\documentclass[review,11pt]{elsarticle}

\usepackage{epsfig}
\usepackage{bm}
\usepackage{lineno,hyperref}
\modulolinenumbers[5]
\usepackage{amsmath,amssymb,amsfonts,amsthm,bm}
\usepackage{cancel}
\usepackage{float}
\usepackage{miller}
\usepackage{subcaption}
\usepackage{booktabs}
\usepackage[margin=1in]{geometry}
\usepackage{algorithm,algorithmic}
\usepackage{textgreek}
\usepackage{soul}
\usepackage{todonotes}
\usepackage{color}
\usepackage{xcolor}
\usepackage{relsize} 
\usepackage{siunitx} 
\usepackage{soul}

\hypersetup{
    colorlinks=true
    }

\definecolor{ic}{HTML}{FF4047}
\definecolor{ea}{HTML}{FF6347}

\biboptions{sort&compress,authoryear}

\makeatletter
\def\ps@pprintTitle{%
  \let\@oddhead\@empty
  \let\@evenhead\@empty
  \let\@oddfoot\@empty
  \let\@evenfoot\@oddfoot
}
\makeatother

\begin{document}
  \hypersetup{
    citecolor=red,
    linkcolor=blue,   
    urlcolor=magenta,
    pdfauthor={Name}
    }



\title{Phase field dislocation dynamics formulation coupled with Fourier based micromechanics solver and its application to grain boundary-dislocation interactions.}

\author[1]{Brayan Murgas\corref{corresp}} \ead{bmurgas@lanl.gov}
\author[2]{Avanish Mishra}
\author[1]{Nithin Mathew}
\author[1]{Abigail Hunter}

\cortext[corresp]{Corresponding author.}

\affiliation[1]{organization={X Computational Physics Division, Los Alamos National Laboratory},
    city={Los Alamos, NM},
    postcode={87545},
    country={USA}}

\affiliation[2]{organization={Theoretical Division, Los Alamos National Laboratory},
    city={Los Alamos, NM},
    postcode={87545},
    country={USA}}

\date{July 2024}

\begin{abstract}

A new phase field dislocation dynamics formulation for homogeneous and heterogeneous materials is presented, which couples micromechanical solvers and the time-dependent Ginzburg-Landau equation. The strain fields are obtained from the micromechanical solver by solving the Lippmann-Schwinger equation, and then used to define energy terms to model the evolution of the dislocations. Grain boundary (GB)-dislocation interactions are studied using the coupled PFDD formulation and by describing GBs as inclusions. GB energy and stiffness tensors are computed from molecular statics simulations and a newly proposed lattice energy term that is dependent on the GB energy is considered in the calculations. Interaction of a screw dislocation with minimum energy and metastable states of low and high angle $\langle$110$\rangle$ symmetric tilt grain boundaries are studied. We show  good agreement between predictions from our phase field dislocation dynamics formulation and molecular dynamics simulations of grain boundary-dislocation interactions.

\end{abstract}

\maketitle

\medskip{}

\noindent \emph{Keywords}: Phase field method, Dislocation dynamics, Molecular Dynamics, Grain boundary, Slip Transfer.

\nolinenumbers

\section{Introduction}
\label{sec:intro}

\let\thefootnote\relax\footnote{LA-UR-25-26159}

New manufacturing processes such as Laser Powder Bed Fusion \citep{gray2017structure, rodriguez2025melt}, Accumulative Roll Bonding (Severe Plastic Deformation) \citep{zhang2023effect}, and  Cold Spray \citep{murgas2024modeling} produce microstructures with complex and dense grain boundary (GB) networks, resulting in material response that varies considerably when compared to microstructures fabricated with classical manufacturing methods and the possibility to tailor properties by controlling the GBs. During deformation, dislocations nucleate, annihilate and move through the grains encountering obstacles that affect their motion. As GBs are inherent in polycrystalline materials, they are a common impediment to dislocation motion and play an important role in controlling the material properties. The Hall-Petch relationship, which predicts that the flow stress is inversely proportional to the square root of the grain size, is the most common link between grain size (or GB density) and a mechanical property \citep{hall1951deformation, petch1953j}. 


However, experimental observations \citep{nieto2024assessment} and molecular dynamics (MD) simulations show that certain GBs do not behave as strong obstacles to dislocation motion \citep{PhysRevMaterials.8.063604}. Rather, GB-dislocation interactions vary, and dislocations can be absorbed, pinned, fully or partially transmitted when impinging on a GB \citep{lim1985continuity, sutton1995interfaces, zaefferer2003influence, priester2012grain}. The type of interaction depends on the type of dislocation, applied thermo-mechanical load and nature of the GB, including the misorientation of the neighboring grains and (atomic or misfit) structure of the GB. Hence, by predicting the GB-dislocation interaction and engineering `favorable' GBs, it is possible to enhance and tailor the in-service properties of the materials, especially during the application of new manufacturing processes where GB networks are dense \citep{gray2017structure, murgas2024modeling, zhang2023effect, rodriguez2025melt}. 

Current criteria for dislocation transmission across GBs are based on experimental observations and consider geometric parameters, stress state and energetics, or combinations of these, in their functional forms. Geometric criteria are based on the alignment of the slip direction $\vec{s}$, plane $\vec{n}$ and intersection line between the slip-plane normal and GB plane normal $\vec{l}$ between incoming (subscript $in$) and outgoing (subscript $out$) slip systems. Examples metrics of different geometric criteria include: $N=(\vec{n}_{in} \cdot\vec{n}_{out})(\vec{s}_{in} \cdot\vec{s}_{out}) + (\vec{n}_{in} \cdot \vec{s}_{out})(\vec{n}_{out} \cdot \vec{s}_{in})$ \citep{livingston1957multiple}, $LRB=(\vec{l}_{in} \cdot \vec{l}_{out})(\vec{s}_{in} \cdot \vec{s}_{out})$ \citep{shen1986dislocation}, $m'=(\vec{n}_{in} \cdot\vec{n}_{out})(\vec{s}_{in} \cdot \vec{s}_{out})$ \citep{luster1995compatibility}, residual Burgers vector $b_r$ \citep{marcinkowski1970dislocation, bollmann2012crystal}, misorientation or disorientation \citep{aust1954effect, clark1954mechanical} and $\lambda$ \citep{werner1990slip}. Stress-based criteria consider the Schmid factor of the incoming and outgoing slip systems \citep{reid2016deformation}, the generalized Schimd factor \citep{bieler2014grain} or the resolved shear stress of the incoming and outgoing slip systems \citep{lee1989prediction}. Combinations of geometric parameters with the accumulated shear or the Schmid factor \citep{bieler2014grain} have been proposed. Lastly, other criteria consider the energy change of the interaction using a line tension model \citep{koning2002modelling}. In this model, the energy of the incoming, outgoing and residual dislocations is considered, as well as GB dislocations and GB energy barriers \citep{tsuru2009fundamental, sangid2011energy, li2020hydrogen, li2022defect}.

Lower scale simulations have provided new insights into GB-dislocations interactions that motivate the need to account for the structure of the GB itself. MD simulations are ideal to study unit processes dictating GB-dislocation interactions over a wide range of GB structures and types of dislocation \citep{spearot2014insights, adams2019atomistic, PhysRevMaterials.8.063604, dang2025dislocation, SURESH2023112085}. In \cite{spearot2014insights} and \cite{adams2019atomistic}, geometric criteria, residual Burgers vector and GB disorientation angle have been identified as indicators of dislocation transmission as observed experimentally; however, these studies were limited to minimum energy GB configurations. In \cite{PhysRevMaterials.8.063604, dang2025dislocation}, GB-dislocation interactions were studied for stable and metastable GB configurations, and different reactions were observed for stable and metastable GB configurations. These results indicate that GB structure, including minimum energy and metastable states, should be considered in the development of multiscale material models. 

Due to time and length scale limitations of atomistic simulations, mesoscale models are necessary to upscale information and new mechanisms to the continuum scale. Examples of mesoscale models are Discrete Dislocation Dynamics (DDD) \citep{li2009strengthening}, Coupled Atomistic/Discrete Dislocation (CADD) \citep{dewald2006multiscale, dewald2007multiscale, dewald2011multiscale, tadmor1996quasicontinuum, shilkrot2002coupled}, and Concurrent Atomistic-Continuum (CAC) \citep{xu2016sequential, su2023multiscale, peng2022atomistic, chen2005atomistic}. In DDD, dislocations are comprised of segments and the precision of simulations highly depends on the rules defined \textit{a priori} \citep{cai2004mobility, bertin2024enhanced}. In \cite{li2009strengthening}, GBs were described as geometric interfaces and the interaction was simplified to permeable or impermeable. In \cite{bamney2021role}, the authors presented a DDD framework in which the structure of stable and metastable GBs were represented using dislocation arrays. However, a transmission rule needs to be defined and, e.g., in \cite{bamney2021role}, the authors used the LRB criterion to model transmission \citep{shen1986dislocation}. Moreover, these studies did not account for partial dislocations and the stacking fault between them, which can impact the GB-dislocation interaction \citep{PhysRevMaterials.8.063604}. 

CADD depends on the definition of defects in the continuum domain and the treatment of the internal boundary force that connects the quasi-continuum and continuum defect problems without the introduction of ghost forces. CADD simulations with GBs have been limited to 2D studies, hence limiting their application to special GBs with specific symmetries \citep{dewald2006multiscale, dewald2007multiscale, dewald2011multiscale}. CAC has been applied to study GB-dislocation interactions in face centered cubic (FCC) metals \citep{xu2016sequential, su2023multiscale} and dislocation pileup in a bicrystal system with square and hexagonal lattices \citep{peng2022atomistic}. CAC simulations are faster than MD simulations, however, as in other coarse-graining methodology (CAC and CADD), the time step of the simulation is imposed by the events on the atomic scale domain. Despite the technical advantages of CAC over CADD, the studies have been limited to 2D or 2.5D domains. On the other hand, CADD and CAC need to reconstruct the GB structure at the atomic scale and are limited to materials with available interatomic potentials.

Another mesoscopic tool available to study GB-dislocation interactions is the phase field dislocation dynamics (PFDD) model \citep{koslowski2002phase, beyerlein2016understanding}. In this model, dislocations evolve through minimization of the total system energy, hence, their interactions and reactions follow an energetically favorable pathway. \cite{lei2013phase} extended the PFDD formulation to heterogeneous materials using a virtual strain formulation to describe the effect of heterogeneities within the total system energy \citep{wang2002phase}. This type of PFDD formulation uses an Eshelby inclusion approach to account for the heterogeneous elastic constants within the system. Thus, the entire system is given the elastic constants of the matrix material and the regions that are meant to be inhomogeneous have a non-zero virtual strain (also sometimes called eigen strain) field to account for the differences. In other words, there are non-zero values of the virtual/eigen strains only in the inclusions, and as the dislocation evolves the nine components of the virtual/eigen strain evolve every time step using an additional kinetic equation. In \cite{lei2013phase}, the evolution of dislocations in a heterogeneous material with voids were considered. In \cite{zeng2016phase}, PFDD was used for the first time to model bimetal interfaces also using the virtual strain formulation. In addition to employing the virtual strains to account for elastic inhomogeneities, a misfit strain was also introduced to describe the lattice mismatch at the bimetal interface. This strain component was derived using linear elasticity and plane stress \citep{hoagland2002strengthening}. Such a misfit strain could also be computed from molecular statics (MS) \citep{kharouji2024atomistic} as well as disclination mechanics using the fast Fourier transform (FFT) \citep{berbenni2018fast} or the finite element method \citep{zhang2018finite}, and use it as an additional eigen strain in the PFDD model. More recently, \cite{ma2022dislocation} studied an incoherent twin boundary (ITB) consisting of an array of three partial dislocations. The ITB material parameters and structure were informed by MS. GB-dislocation interaction events were in good agreement between PFDD and MD. This methodology allows the method to describe GB structures in detail and study GB-dislocation interactions precisely; however, is limited to certain systems where the GB structure can be defined in terms of well-defined dislocation arrays.

Clearly, FFT-based models have been able to successfully model a diverse set of problems \citep{moulinec1994fast, lebensohn2001n, bertin2015fft, berbenni2014numerical, michel2001computational, chen2019fft, chen2015integrated, vidyasagar2017predicting} and carries advantages such as a computational complexity of $O(N log N)$ compared to $O(N^2)$ of the finite element method, reduced memory allocation needs (matrix free formulation), and the possibility to use images as inputs as pixels (voxels) are used in 2D (3D). Thus, we first present a new PFDD formulation for heterogeneous materials coupled to a FFT-based micromechanical solver. The coupled formulation we present is consistent with the homogeneous PFDD formulation \citep{koslowski2002phase}; however, compared to the virtual strain formulation \citep{lei2013phase}, it does not require virtual strain fields to evolve the strain within the inclusions. The results presented here use the non-linear FFT micromechanical solver known as the basic scheme introduced by \cite{moulinec1994fast} with an eigen strain that represents the plastic deformation from the dislocation. However, different solvers could be used if desired. Note that the basic scheme requires a lot of iterations to converge if the ratio of the elastic properties tend to infinity or zero. Second, we present simulations of GB-dislocation interactions using an approach that can account for general GB structures, removing the limitation to well-defined GB structures where the structure must be determined \textit{a priori}. This approach is demonstrated through simulations of bicrystals where the GB is defined as an inclusion. The simulations parameters are informed by MS \citep{PhysRevMaterials.8.123605}, and compared to recent MD simulations \citep{dang2025dislocation}. In particular, we focus on the observations made by \cite{dang2025dislocation}: GB-dislocation interaction varies depending on the applied stress, misorientation, and whether it is a minimum energy or metastable GB. MD simulations show that interactions can vary and involve only the leading partial during transmission (partial transmission). To capture this, PFDD simulations were performed using partial dislocations \citep{hunter2013dependence}. In Section~\ref{sec:methods}, the PFDD formulation is presented. This section describes how the micromechanical problem is solved, and how the total strain is used in the elastic energy terms that describe the evolution of the dislocations. Next, Section~\ref{sec:results} presents the PFDD simulation results of dislocation reactions with different GBs in Cu. Finally, concluding remarks are given in Section~\ref{sec:conclusion}.

\section{Phase field dislocation dynamics formulation coupled with the Lippmann-Schwinger equation}
\label{sec:methods}

\begin{figure}[ht]
   \centering
   \includegraphics[width=0.35\textwidth]{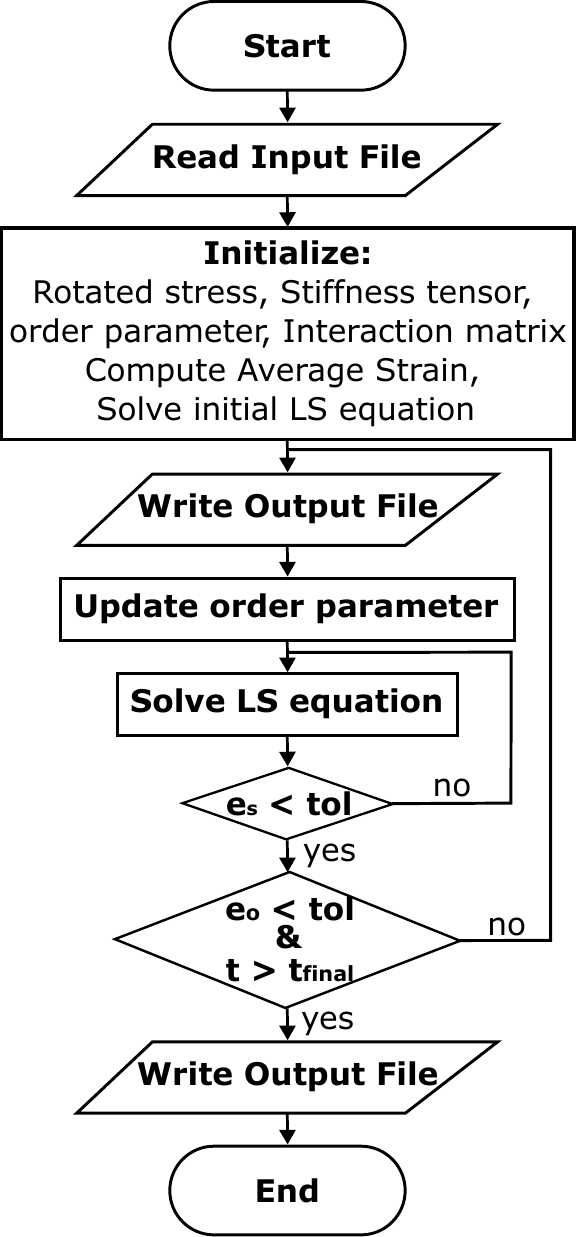}
   \caption{Pseudo-algorithm of the new PFDD formulation. $e_s=1\times 10^{-5}$ and $e_o=1\times 10^{-8}$ are stress and order parameter convergence tolerances.}
   \label{fig:flowchartcoupled}
\end{figure}

In the PFDD formulation, dislocations are described using a scalar order parameter, $\zeta^{\alpha}$, that is defined on each active slip system, $\alpha$. $\zeta^{\alpha}=0$ and $\zeta^{\alpha}=1$ correspond to a perfect crystal and a crystal that has undergone slip, respectively, while non-integer values between 0 and 1 represent a distorted crystal structure (i.e., the location of the dislocation core). This coupled formulation uses two different sets of equations, one to obtain the total strain through solving the Lippmann-Schwinger (LS) equation and the other to obtain the evolution of dislocations in the system following the pseudo-algorithm shown in Figure~\ref{fig:flowchartcoupled}. Note that every time step the staggered scheme follows this sequence: (i) the mechanical problem is solved, (ii) the energy terms are defined, and (iii) the dislocation evolves. The three different components of the algorithm are described in the subsections below.

\subsection{Dislocation mechanics solving the Lippmann-Schwinger equation}

The stress of the heterogeneous domain at a point $x$ is computed as
\begin{equation}
    \sigma_{ij} = C_{ijkl}(x) \varepsilon_{kl}^e,
\end{equation}
where $C_{ijkl}(x)$ is the fourth order stiffness tensor of the phase or material at point $x$ and $\varepsilon^e$ is the elastic strain. The elastic strain is decomposed as
\begin{equation}
    \varepsilon^e_{ij} = \varepsilon_{ij} + E_{ij} - \varepsilon^p_{ij},
    \label{eqn:straindecomposition}
\end{equation}
where the total strain is split into a fluctuation term $\varepsilon$ and a prescribed volume average term $E$ corresponding to the homogeneous strain, and $\varepsilon^p$ is the plastic strain. The plastic strain depends on the order parameter as \citep{koslowski2002phase, beyerlein2016understanding, nabarro1951cxxii}
\begin{equation}
    \varepsilon^p_{ij} = \dfrac{1}{2} \sum_{\alpha=1}^N \dfrac{ b \zeta^{\alpha} }{ d } ( s^{\alpha}_i n^{\alpha}_j + s^{\alpha}_j n^{\alpha}_i ),
\end{equation}
where $b$ is the Burgers vector magnitude, $d$ is the interplanar spacing, $N$ is the number of active slip systems, and, $s^{\alpha}$ and $n^{\alpha}$ are the slip direction and slip plane normal of slip system $\alpha$. The sum over $\alpha$ means that the plastic strain is proportional to the number of active slip systems. In this formulation, all of the order parameters are aligned with perfect dislocation slip systems in FCC metals, as this is the material system studied later in the Results Section (Section~\ref{sec:results}).  Partial dislocations are modeled through linear combinations of the 3 perfect dislocation order parameters on a given slip plane. We note that the PFDD method is not limited to FCC systems, but has also been applied to body centered cubic (BCC) \citep{fey2024role, fey2022phase}, hexagonal close packed (HCP) \citep{albrecht2021asymmetric}, and semiconductor systems \citep{kim2022phase}. For the case of HCP materials, the Burgers vector magnitude and interplanar distance vary with the active slip systems, and would require that $b$ and $d$ both also be a function of the slip system, $\alpha$. However, this is not the case for the FCC system considered here, and these superscripts are dropped for convenience.

Introducing a reference material with stiffness $C^0$ and using the decomposition of the elastic strain, the stress may be redefined as
\begin{equation}
    \begin{split}
    \sigma_{ij} & = ( C_{ijkl}^0 + \Delta C_{ijkl} (x) ) ( \varepsilon_{kl} + E_{kl} - \varepsilon^p_{kl} ) \\
    & =  C_{ijkl}^0  ( \varepsilon_{kl} + E_{kl} ) + \Delta  C_{ijkl}(x)  ( \varepsilon_{kl} + E_{kl} ) - C_{ijkl}(x) \varepsilon^p_{kl}.
    \end{split}
    \label{eqn:stress}
\end{equation}
The last two terms in the right hand side of Equation~\ref{eqn:stress} form the polarization tensor, defined as
\begin{equation}
    \tau_{ij} = \Delta  C_{ijkl}(x)  ( \varepsilon_{kl} + E_{kl} ) - C_{ijkl}(x) \varepsilon^p_{kl},
\end{equation}
or more generally
\begin{equation}
    \tau_{ij} = \sigma_{ij} - C_{ijkl}^0  ( \varepsilon_{kl} + E_{kl} ).
\end{equation}
Using the polarization stress and Equation~\ref{eqn:stress}, the balance of linear momentum gives the following heterogeneous partial differential equation
\begin{equation}
    (C_{ijkl}^0 \varepsilon_{kl}  )_{,j} = -( \tau_{ij}(\varepsilon) )_{,j}.
\end{equation}
Introducing the Green's function, the above equation can be solved in the form of an integral Lippmann-Schwinger equation for the unknown strain \citep{kroner1971statistical}:
\begin{equation}
    \varepsilon_{ij} = - (\Gamma^0_{ijkl} \star \tau_{kl} (\varepsilon) )(x),
    \label{eqn:LS}
\end{equation}
where $\star$ denotes spatial convolution and $\Gamma_{ijkl}^0$ is the modified Green's tensor defined as the second derivative of the Green's function $\Gamma_{ijkl}^0 = - G^0_{ki,jl}$ $\forall \xi \ne 0$ and $\hat{\varepsilon}_{ij}(0) = 0$ if $\xi=0$. In this work, Equation~\ref{eqn:LS} is solved using the basic scheme proposed by \cite{moulinec1994fast} and reference stiffness tensor $C^0_{ijkl} = 0.5 ( C^{min}_{ijkl} + C^{max}_{ijkl} )$, where $C^{min}_{ijkl}$ and $C^{max}_{ijkl}$ are the minimum and maximum elastic properties, see Algorithm~\ref{alg:basic} in the Supplementary Material.

\subsection{Dislocation evolution using the time-dependent Ginzburg-Landau (TDGL) kinetic equation} 

The total energy is composed of three contributions \citep{koslowski2002phase, beyerlein2016understanding}:
\begin{equation}
    \Psi = \Psi_{elas} + \Psi_{latt} - \Psi_{ext},
    \label{eqn:totalenergy}
\end{equation}
where $\Psi_{elas}$, $\Psi_{latt}$ and $\Psi_{ext}$ are the elastic, lattice and external energy. Note that in some PFDD formulations there is an additional gradient energy term representing the excess energy generated by the non-uniformity of the dislocation \citep{xu2019comparison, shen2003phase, kim2021phase}, which is neglected in the current work. The elastic energy accounts for the elastic interactions between dislocations and is defined as
\begin{equation}
    \Psi_{elas} = \frac{1}{2} \int_V C_{ijkl} (x) \varepsilon^e_{ij} \varepsilon^e_{kl} dV =  \frac{1}{2} \int_V C_{ijkl} (x) ( E_{ij} + \varepsilon_{ij} - \varepsilon^p_{ij} ) ( E_{kl} + \varepsilon_{kl} - \varepsilon^p_{kl} ) dV,
    \label{eqn:elastreal}
\end{equation}
with $V$ being the volume of the simulation domain. To define the elastic energy of the heterogeneous system, the elastic strain can be derived from the decomposition of the total strain in Equation~\ref{eqn:straindecomposition} and the definition of the total strain in Equation~\ref{eqn:LS}. Instead of applying the Plancherel theorem \citep{plancherel1910contribution} to obtain the elastic energy in the Fourier space, in this work we compute the elastic energy in the real space using Equation~\ref{eqn:elastreal}.

The lattice energy represents the energy required to move the dislocation core through the crystal lattice by the breaking and reforming of atomic bonds. It can be expressed using different forms such as piecewise quadratic functions \citep{koslowski2002phase}, sinusoidal functions \citep{xu2020frank, peng20203d} or Fourier series \citep{PhysRevB.84.144108, schoeck2001core, shen2004incorporation}, depending on the simulation parameters:
\begin{equation}
    \Psi_{latt} = \sum_{\alpha=1}^N \int \phi(\zeta^{\alpha}) dV
    \label{eqn:lattenergy}
\end{equation}
where $\phi( \zeta^{\alpha})$ are the periodic potentials. For simulations in this work, it is necessary to account for partial dislocations. For this, we consider the $(111)$ slip plane and three different slip directions with order parameters, $\zeta^{\alpha}_1$, $\zeta^{\alpha}_2$ and $\zeta^{\alpha}_3$, that correspond to the directions $s_1 = \frac{\sqrt{2}}{2}[0 \bar{1} 1]$, $s_2 = \frac{\sqrt{2}}{2}[1 0 \bar{1}]$ and $s_3 = \frac{\sqrt{2}}{2}[\bar{1} 1 0]$, respectively \citep{hunter2013dependence}. 

The lattice energy in Equation~\ref{eqn:lattenergy} can then be redefined as 
\begin{equation}
    \Psi_{latt} = \sum_{\alpha=1}^N \int \phi(\zeta^{\alpha}_1,\zeta^{\alpha}_2,\zeta^{\alpha}_3) dV,
    \label{eqn:lattenergypartials}
\end{equation}
where $\phi(\zeta^{\alpha}_1,\zeta^{\alpha}_2,\zeta^{\alpha}_3)$ is implemented as a Fourier series \citep{PhysRevB.84.144108, schoeck2001core, shen2004incorporation}:
\begin{equation}
    \begin{split}
    \phi(\zeta_1,\zeta_2,\zeta_3) & = c_0 + c_1 [ \text{cos} \ 2 \pi (\zeta_1 - \zeta_2) + \text{cos} \ 2 \pi (\zeta_2 - \zeta_3) + \text{cos} \ 2 \pi (\zeta_3 - \zeta_1) ] \\
    & + c_2 [ \text{cos} \ 2 \pi (2 \zeta_1 - \zeta_2 - \zeta_3) + \text{cos} \ 2 \pi (2 \zeta_2 - \zeta_3 - \zeta_1) + \text{cos} \ 2 \pi (2 \zeta_3 - \zeta_1 - \zeta_2) ]  \\
    & + c_3 [ \text{cos} \ 4 \pi (\zeta_1 - \zeta_2) + \text{cos} \ 4 \pi (\zeta_2 - \zeta_3) + \text{cos} \ 4 \pi (\zeta_3 - \zeta_1) ] \\
    & + c_4 [ \text{cos} \ 4 \pi (3 \zeta_1 - \zeta_2 - 2 \zeta_3) + \text{cos} \ 4 \pi (3 \zeta_1 - 2 \zeta_2 - \zeta_3) \ \text{cos} \ 4 \pi (3 \zeta_2 - \zeta_3 - 2 \zeta_1) \\
    & + \text{cos} \ 4 \pi (3 \zeta_2 - 2 \zeta_3 - \zeta_1) \ \text{cos} \ 4 \pi (3 \zeta_3 - \zeta_1 - 2 \zeta_2) + \text{cos} \ 4 \pi (3 \zeta_3 - 2 \zeta_1 - \zeta_2) ]  \\
    & + a_1 [ \text{sin} \ 2 \pi (\zeta_1 - \zeta_2) + \text{sin} \ 2 \pi (\zeta_2 - \zeta_3) + \text{sin} \ 2 \pi (\zeta_3 - \zeta_1) ] \\
    & + a_3 [ \text{sin} \ 4 \pi (\zeta_1 - \zeta_2) + \text{sin} \ 4 \pi (\zeta_2 - \zeta_3) + \text{sin} \ 4 \pi (\zeta_3 - \zeta_1) ].
    \end{split}
    \label{eqn:lattenergyFourier}
\end{equation}
The expansion coefficients $c_{0-4}$ and $a_{1,3}$ can be obtained by fitting the $\gamma$-surface computed using MD or DFT; in this work we used the same coefficients computed by \cite{ma2022dislocation}.

The external energy accounts for interactions between the applied stress and dislocations and is defined as
\begin{equation}
    \Psi_{ext} = \sigma^{app}_{ij} E_{ij} V.
\end{equation}
where $\sigma^{app}$ is the applied stress. The homogeneous strain is determined by the boundary conditions. For a system in equilibrium, the homogeneous strain is determined by minimizing the total strain energy $\Psi_{str} = \Psi_{elas} - \Psi_{ext}$ with respect to the homogeneous strain
\begin{equation}
    \dfrac{\partial \Psi_{str}}{\partial E_{ij}} = 0.
\end{equation}
Hence, the homogeneous strain is 
\begin{equation}
    E_{kl} =  \big< S_{ijkl} \big> \left( \sigma_{ij}^{app} + \big< \sigma_{ij}^{p}  \big> - \big< \sigma_{ij} \big>  \right),
    \label{eqn:homstraincoupled}
\end{equation}
where $ \big< S_{ijkl} \big> = \big< C_{ijkl} \big>^{-1} $, $ \big< C_{ijkl} \big> =  \frac{1}{V} \int_v C_{ijkl} (x) dV $, $\big< \sigma_{ij}^{p}  \big> = \frac{1}{V} \int_v C_{ijkl} (x) \varepsilon_{kl}^p dV$, and $ \big< \sigma_{ij} \big> =  \frac{1}{V} \int_v \Delta C_{ijkl} (x) \varepsilon_{kl} dV $. For an applied strain $E=\varepsilon_{ij}^{app} $,the external energy is defined as $\Psi_{ext} = C_{ijkl}(x) \varepsilon_{kl}^{app} \varepsilon_{ij}^{app} V$. In a homogeneous system, Equation~\ref{eqn:homstraincoupled} is redefined as $E_{kl} = \frac{1}{V} \int_v \varepsilon_{kl}^p dV + S_{klij} \sigma_{ij}^{app}$ which simplifies to the definition of the external energy defined in the homogeneous PFDD formulation \citep{jin2001phase, koslowski2002phase}.

Finally, the dislocation order parameter evolves using a time-dependent Ginzburg-Landau (TDGL) kinetic equation, which minimizes the total energy
\begin{equation}
    \dfrac{ \partial \zeta^{\alpha} }{ \partial t } = - L \dfrac{ \delta \psi }{ \delta \zeta^{\alpha} },
\end{equation}
where $L$ is a coefficient that represents the mobility of the dislocations, $\psi$ is the energy density and a variational derivative is used. The variational derivative can be transformed into a partial derivative as \citep{giaquinta2013calculus} (c.f. Chapter 1, Section 2.1)
\begin{equation}
    \dfrac{ \partial \zeta^{\alpha} }{ \partial t } = L \left( \nabla \cdot \dfrac{ \partial \psi }{ \partial \nabla \zeta^{\alpha} } - \dfrac{ \partial \psi }{ \partial \zeta^{\alpha} } \right),
\end{equation}
as the gradient term is neglected in this work, the TDGL equation is
\begin{equation}
    \dfrac{ \partial \zeta^{\alpha} }{ \partial t } = - L  \dfrac{ \partial \psi }{ \partial \zeta^{\alpha} } \, .
\end{equation}
This equation is discretized in time following the forward Euler method as
\begin{equation}
    \zeta^{\alpha}_{t+1} = \zeta^{\alpha}_{t} - L \Delta t \dfrac{ \partial \psi }{ \partial \zeta^{\alpha}_{t} } \, ,
\end{equation}
where the subscripts $t$ and $t+1$ refers to the current and next time step. More details about the formulation are shown in Supplementary Section~\ref{suppSec:details}. 

\subsection{Representation of grain boundaries}

\begin{figure}[ht]
   \centering\leavevmode
   \includegraphics[width=90mm]{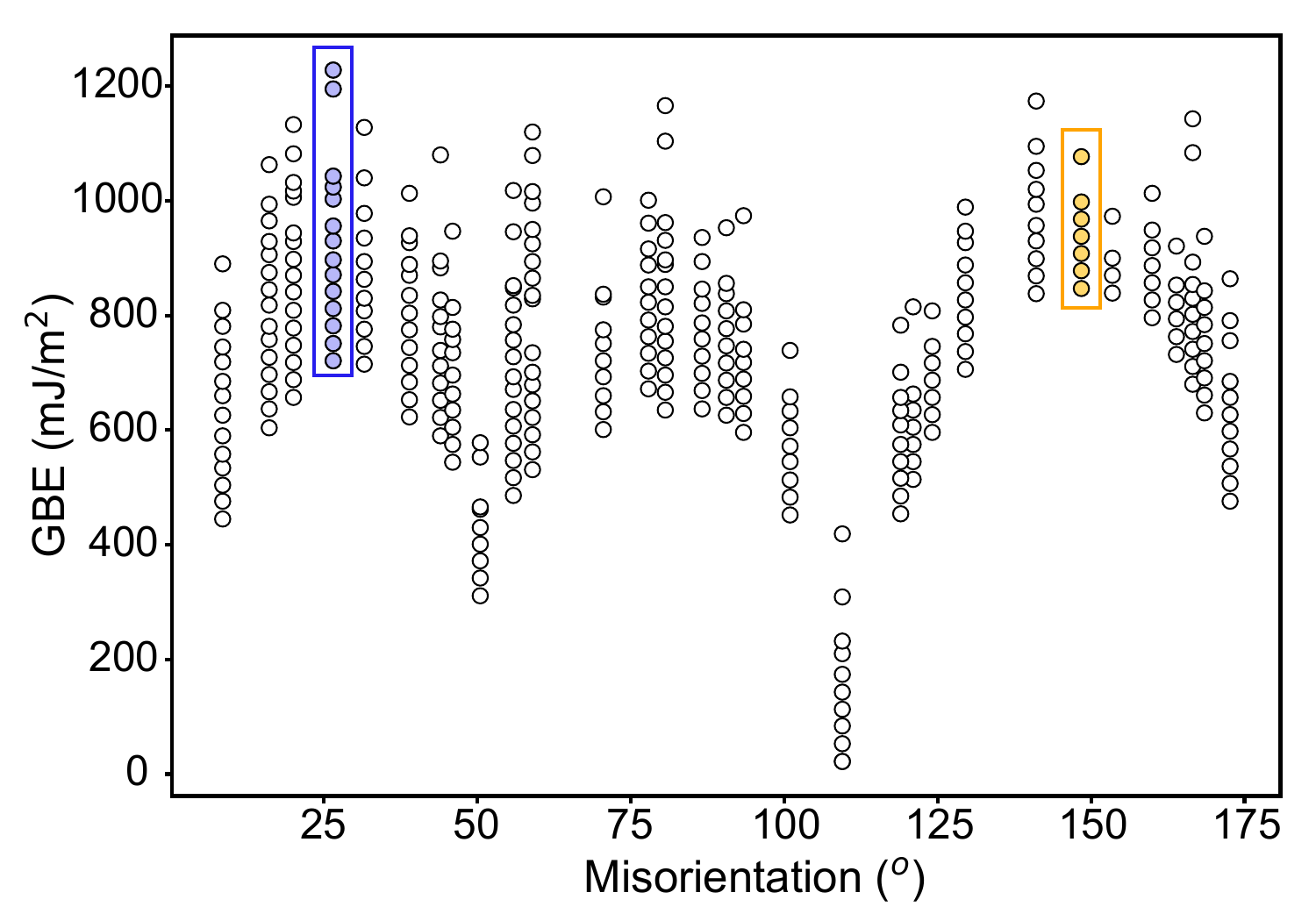}
   \caption{Grain boundary energy (GBE) as a function of misorientation angle for $<110>$ symmetric tilt grain boundary. The GBE of the LAGB and HAGB are colored in blue and orange, respectively. The same data points were used to study GB-dislocation interactions by \cite{dang2025dislocation}.}
   \label{fig:gb-energy}
\end{figure}

In previous work \citep{dang2025dislocation}, dislocation-GB interactions were studied with MD for several different symmetric tilt GBs in Cu including both stable and metastable GB configurations. From this database of GB structures \citep{dang2025dislocation}, we selected two $\langle$110$\rangle$ symmetric tilt GBs and their respective metastable states to consider in this paper. The first GB is a low to medium misorientation angle GB (LAGB) with (1,1,6) GB plane and a misorientation angle of 26.5$^\circ$ and the second GB is a high misorientation angle GB (HAGB) with (5,5,2) GB plane and a misorientation angle of 148.4$^\circ$. In this work, GBs are represented as inclusions with different properties compared to the surrounding grains. We use MS calculations to define three GB properties: (1) width (2) second-order elastic constants (SOECs), and (3) GB energy (GBE). The width of the GB was determined using a two-class Gaussian Mixture Model clustering of the atomistic configuration as explained by \cite{PhysRevMaterials.8.123605}. Figure~\ref{fig:gb-energy} shows the $<110>$ symmetric tilt GBE as a function of the misorientation angle, highlighting the GBs selected for this work in boxes. The minimum energy or stable state correspond to the lowest GBE value for a given misorientation angle, while the metastable states have higher GBE values. The GBE of the LAGB and HAGB are colored in blue and orange in Figure~\ref{fig:gb-energy}. 

\begin{figure}[ht]
   \centering\leavevmode
   \includegraphics[width=160mm]{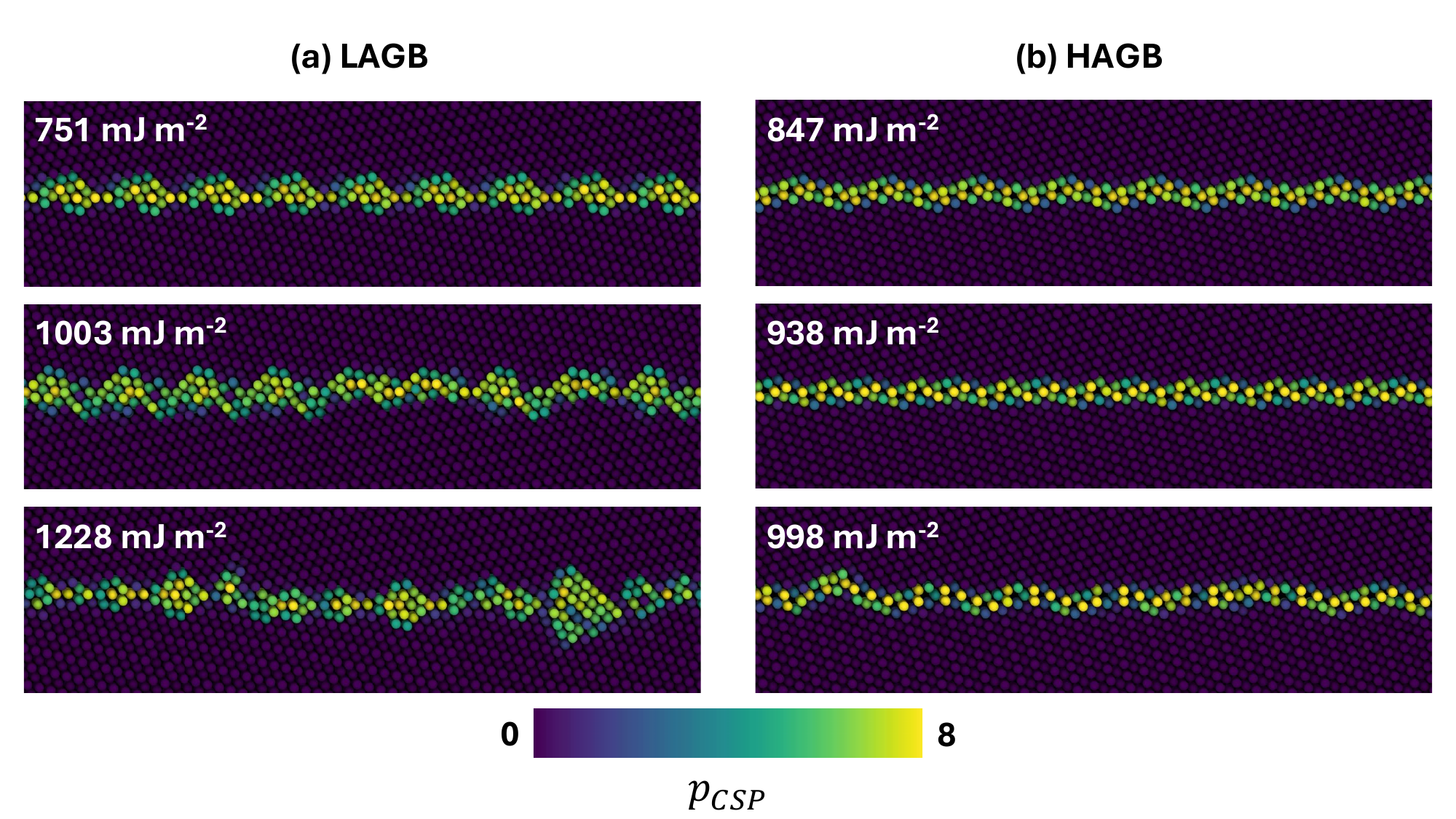}
   \caption{Grain boundary structure of $\langle 110 \rangle$ symmetric tilt. (a) Low-angle GB (LAGB) with the (1,1,6) GB plane and (b) a high-angle GB (HAGB) with the (5,5,2) GB plane in Cu. The minimum energy structures are shown in the top (751 and 847 mJ m$^{-2}$ for the LAGB and HAGB, respectively) and two metastable structures are shown in the middle (1003 and 938 mJ m$^{-2}$ for the LAGB and HAGB, respectively) and bottom (1228 and 998 mJ m$^{-2}$ for the LAGB and HAGB, respectively). The color bar corresponds to the centrosymmetry parameter computed using Ovito.}
   \label{fig:gb-width}
\end{figure}

\begin{figure}[ht]
   \centering\leavevmode
   \includegraphics[width=120mm]{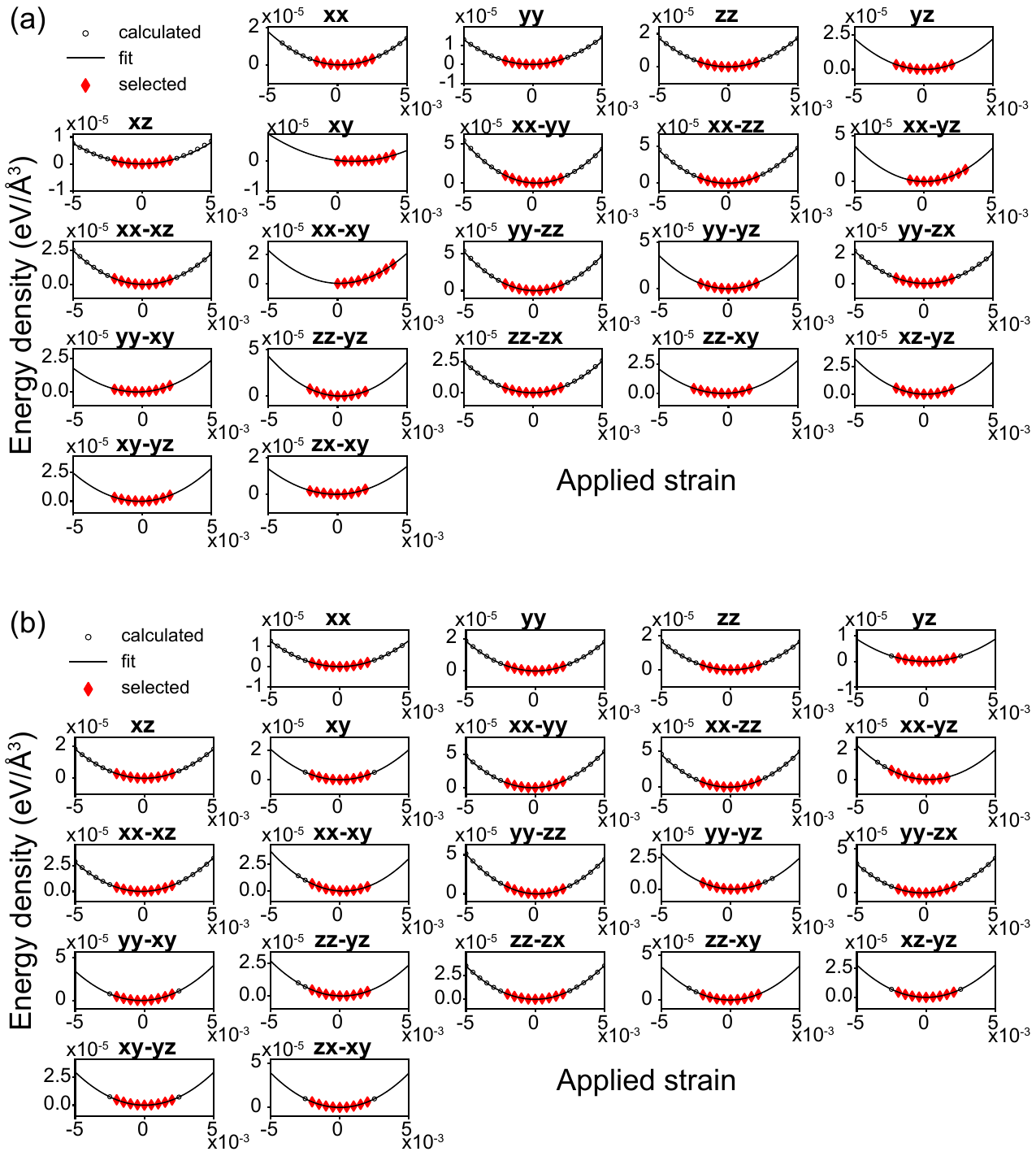}
   \caption{Energy density versus applied strain for the grain-boundary (GB) region in minimum-GB energy structures from MS: (a) a low-angle GB (LAGB) with the (1,1,6) GB plane and (b) a high-angle GB (HAGB) with the (5,5,2) GB plane. The nine lowest-energy points (red diamonds) are used for quadratic fitting to calculate the stiffness tensor. The title of each subplot mentions the direction of applied strain.}
   \label{fig:energy-strain}
\end{figure}

The atomistic structure of GBs used in this work are shown in Figure \ref{fig:gb-width}; the same GBs and interatomic potential were used by \cite{dang2025dislocation}. The atoms in Figure \ref{fig:gb-width} are colored by the Centro Symmetry Parameter (CSP) \cite{kelchner1998dislocation} computed using OVITO \cite{OVITO}. The minimum energy GBs have well-ordered atomic arrangements. Significant deviations from these are clearly present in the higher energy metastable GBs with different microscopic degrees of freedom. To determine the SOECs of the GB region, we performed affine deformation of a bicrystal with 3D periodic boundary conditions. A prescribed set of lattice strains ($\pm0.2\%$) was applied to an energy minimized bi-crystal under the assumption of general elastic anisotropy. Per-atom energies for the atoms in the GB region were used to calculate the energy density in the GB region as a function of the applied strain. The elastic stiffness coefficients were determined from these energy density-strain data. We assume that the strain in the bi-crystal is homogeneous and equal to the applied strain. The Lam\'e constants are determined from the SOECs using the Voigt average, see Figure~\ref{fig:energy-strain}.

We perturb the lattice energy in the GB region using an additional term that accounts for the difference in atomic structure and its effect on the evolution of the dislocation core. This additional energy barrier is relevant to the resulting dislocation-GB interaction; however, it cannot be determined analytically but can be estimated from atomic scale simulations. Previous studies have estimated the energy barrier using nudged elastic band (NEB) method \citep{tsuru2009fundamental} and MD simulations \citep{sangid2011energy, li2020hydrogen, li2022defect}. In \cite{sangid2011energy} and \cite{li2020hydrogen}, the energy barrier and GBE are correlated using a power law equation; however, the number of data points used are not enough to create a robust model. Thus, in this work, we take a different approach and use the GBE to scale the lattice energy in the bulk grains (see Equation~\ref{eqn:lattenergypartials}) to define the perturbation on the lattice energy in the GB region as
\begin{equation}
     \Psi^{\mathsmaller{GB}}_{latt} = \Psi_{latt} C_{\mathsmaller{GB}} \dfrac{\gamma_{\mathsmaller{GB}} - \gamma_{usf}}{ \gamma_{usf} } \dfrac{ \mu_{grain} }{  \mu_{\mathsmaller{GB}}} \, ,
     \label{eqn:phigb}
\end{equation}
where $C_{\mathsmaller{GB}}$ is a fitting constant, $\gamma_{\mathsmaller{GB}}$ is the GBE informed by MS calculations (See Figure~\ref{fig:gb-energy}), $\gamma_{usf}$ is the unstable stacking fault energy, $\mu_{grain}$ is the shear modulus of the grain and $\mu_{\mathsmaller{GB}}$ is the shear modulus of the GB. This term acts as a perturbation of the $\gamma$-surface in the bulk grain to account for the disordered nature of the GB with respect to the bulk crystal so the total energy at the GB region is calculated as $\Psi = \Psi_{elas} + \Psi_{latt} + \Psi^{\mathsmaller{GB}}_{latt} - \Psi_{ext}$. To align with previous work, this term could be modified as $\Psi^{\mathsmaller{GB}}_{latt} = C_{\mathsmaller{GB}} \gamma_{\mathsmaller{GB}}^{-a}, $ where $C_{\mathsmaller{GB}}$ and $a$ are fitting parameters \citep{sangid2011energy, li2020hydrogen}. We highlight again that the energy barrier to dislocation glide within a GB due to the atomic disorder is still an open research question, and is expected to have strong dependence on the GB structure. Thus, these equations are representative of this effect, but may need to be redefined for GB structures not considered in this paper. In particular, specialized GB structures such as cube-on-cube orientation or coherent twin boundaries with low values of GBE $\gamma_{\mathsmaller{GB}}$, are likely examples of cases in which different descriptions of this energy barrier may be necessary.

The different GB widths, GB energies, elastic constants and GB-dislocation reactions as predicted with MD are shown in Table~\ref{tab:GBprops}. The Lam\'e constants of the Cu grains surrounding the GB region are $\mu_{grain}=40.43\text{ GPa}$ and $\lambda_{grain}=111.12\text{ GPa}$ and $\gamma_{usf} = 163.7$ mJ/m$^2$. For each GB, the minimum energy structure plus two higher metastable structures were selected for this study. We note that we were unable to determine the SOECs for some metastable GBs as they were unstable under some deformation modes. Therefore, the Lam\'e constants and GB width of the minimum energy GB structure is used also for the metastable cases. 

\begin{table}[ht]
    \centering
    \begin{tabular}{c|ccc|ccc}
    \toprule
         Type & GB Energy & $\mu,\lambda$ & Width &  & MD Reaction &   \\
              & mJ/m$^2$ & GPa & \AA & 250 MPa & 500 MPa & 750 MPa  \\
    \midrule
         LAGB & 751   & 29.16, 119.78 & 18.69 & A/P  & T    &  T    \\
         LAGB & 1003  & 29.16, 119.78 & 18.69 & A/P  & T    &  T    \\
         LAGB & 1228  & 29.16, 119.78 & 18.69 & A/P  & A/P  &  T   \\
         \midrule
         HAGB & 847   & 40.11, 112.11 & 12.6 & A/P  & A/P  &  T   \\
         HAGB & 938   & 40.11, 112.11 & 12.6 & A/P  & A/P  &  T   \\
         HAGB & 998   & 40.11, 112.11 & 12.6 & A/P  & A/P  &  A/P \\
    \bottomrule
    \end{tabular}
    \caption{Grain boundary (GB) properties and respective GB-dislocation reaction at different applied stresses: 250, 500 and 750 MPa \citep{dang2025dislocation}. $\mu$ and $\lambda$ are the Lam\'e constants. A/P and T stand for Absorption or Pinning and Transmission, respectively.}
    \label{tab:GBprops}
\end{table}

\subsection{Material properties and simulation setup}

\begin{figure}[ht]
   \centering\leavevmode
   \includegraphics[width=80mm]{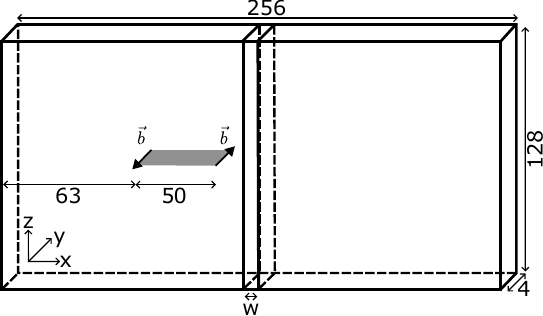}
   \caption{Simulation setup showing the position of the dislocation dipole (in number of grid points) and inclusion representing the GB in a bicrystal. The domain size is 256x4x128. The dislocation dipole is centered in the vertical direction, z, the GB is centered in the horizontal direction, x, and the width of the LAGB and HAGB is 7 and 5 grid points, respectively.}
   \label{fig:HetDisloc}
\end{figure}

A domain of 256x4x128 is used with a grid size that corresponds to the Burgers vector norm of a screw dislocation in Cu ($b=2.556$ \AA). Periodic boundary conditions are used in the three directions and the domain size is chosen such that the periodic interactions are minimized, when bigger domains were used the results did not change. An inclusion, representing the GB region, is located in the center of the domain and the position of the initial dislocation dipole is presented in Figure~\ref{fig:HetDisloc}. A screw dislocation dipole is placed in the left grain, with the left monopole placed at 63 grid points from the domain boundary with a distance of 50 grid points between the dislocations. The domain is subjected to three different applied shear stresses $\sigma_{zy}^{app}=250,500,750$ MPa, resulting in a different reaction for a given GBE value as shown in Table~\ref{tab:GBprops}. The dislocation mobility $L$ is set to $0.2 (\mu_{grain} \Delta t)^{-1}$ with a dimensionless time step of $\Delta t = 1$.

\section{Results and discussion}
\label{sec:results}

The reactions in Table~\ref{tab:GBprops} are reproduced using the PFDD model by setting $C_{\mathsmaller{GB}}$ for each type of GB. The values used in the following simulations for the LAGB and HAGB are 0.04 and 0.156, respectively. As the elastic constants and energies are fixed for a given GB, Equation~\ref{eqn:phigb} may be redefined as
\begin{equation}
     \Psi^{\mathsmaller{GB}}_{latt} = \Psi_{latt} f_{\mathsmaller{GB}},
\end{equation}
with 
\begin{equation}
     f_{\mathsmaller{GB}} = C_{\mathsmaller{GB}} \dfrac{\gamma_{\mathsmaller{GB}} - \gamma_{usf}}{ \gamma_{usf} } \dfrac{ \mu_{grain} }{  \mu_{\mathsmaller{GB}}}.
\end{equation}
The value of $f_{\mathsmaller{GB}}$ is shown in Table~\ref{tab:GBfactor} for each GB. Based on the reactions at different applied stress, the six GBs can be divided into three categories as shown in Table~\ref{tab:GBfactor}. One can notice the value of $f_{\mathsmaller{GB}}$ is close within each category. If $f_{\mathsmaller{GB}}$ is between 0.1989 and 0.284, the GB-dislocation reactions are (A/P, T, T), if $f_{\mathsmaller{GB}}$ is between 0.656 and 0.743, the GB-dislocation reactions are (A/P, A/P, T), and if $f_{\mathsmaller{GB}}$ is equal or higher than 0.801, the GB-dislocation reactions are (A/P, A/P, A/P). The $f_{\mathsmaller{GB}}$ values roughly correlate with strain-gradient metrics from the strain functional descriptors used in the two-class Gaussian Mixture Model clustering in \citep{PhysRevMaterials.8.123605}. More details are provided in the Supplementary Materials, see Supplementary Figures~\ref{fig:P3I0-smallGB}-\ref{fig:P3I4-largeGB}.

\begin{table}[ht]
    \centering
    \begin{tabular}{c|ccc|ccc}
    \toprule
         Type & GB Energy & $C_{\mathsmaller{GB}}$ & $f_{\mathsmaller{GB}}$ &  & PFDD Reaction &  \\
              & mJ/m$^2$ &  &  & 250 MPa & 500 MPa & 750 MPa  \\
    \midrule
         LAGB & 751   & 0.04  & 0.1989  & A/P  & T      &  T     \\
         LAGB & 1003  & 0.04  & 0.284  & A/P  & T (C1) &  T     \\
         \midrule
         LAGB & 1228  & 0.04  & 0.3605  & A/P  & A/P (C2)  &  T     \\
         HAGB & 847   & 0.156 & 0.656  & A/P  & A/P (C3)  &  T (C4)     \\
         HAGB & 938   & 0.156 & 0.743  & A/P  & A/P  &  T     \\
         \midrule
         HAGB & 998   & 0.156 & 0.801  & A/P  & A/P  &  A/P   \\
    \bottomrule
    \end{tabular}
    \caption{Grain boundary energy, parameters ($C_{\mathsmaller{GB}}$ and $f_{\mathsmaller{GB}}$) and resulting PFDD reactions at different applied stresses: 250, 500 and 750 MPa. C1, C2, C3 and C4 are the cases analyzed in detail in the main text.}
    \label{tab:GBfactor}
\end{table}

Among the cases shown in Table~\ref{tab:GBprops}, four different cases are analyzed here in more detail: (C1) LAGB with GBE $\gamma_{\mathsmaller{GB}}=1003$ mJ/m$^2$ under an applied stress of $\sigma_{yz}^{app} = 500$ MPa, (C2) LAGB with GBE $\gamma_{\mathsmaller{GB}}=1228$ mJ/m$^2$ under an applied stress of $\sigma_{yz}^{app} = 500$ MPa, (C3) HAGB with GBE $\gamma_{\mathsmaller{GB}}=847$ mJ/m$^2$ under an applied stress of $\sigma_{yz}^{app} = 500$ MPa and (C4) HAGB with GBE $\gamma_{\mathsmaller{GB}}=847$ mJ/m$^2$ under an applied stress of $\sigma_{yz}^{app} = 750$ MPa. Note that the other cases have been simulated and the GB-dislocation reactions from MD are reproduced, see Supplementary Figures~\ref{fig:disevol-lagb-751-250}-\ref{fig:disevol-hagb-998-750}. C1, C2 and C3 have different energies and same applied stress while C3 and C4 are the same GB with  different applied stresses. $f_{\mathsmaller{GB}}$ for C1, C2 and C3/C4 is 0.284, 0.386 and 0.656, respectively (see Table~\ref{tab:GBfactor}). To the authors knowledge, there is no way to compute analytically the values of $C_{\mathsmaller{GB}}$, for the work presented here the values are chosen to reproduce the MD reactions.

\begin{figure}[ht!]
   \centering\leavevmode
   \includegraphics[width=135mm]{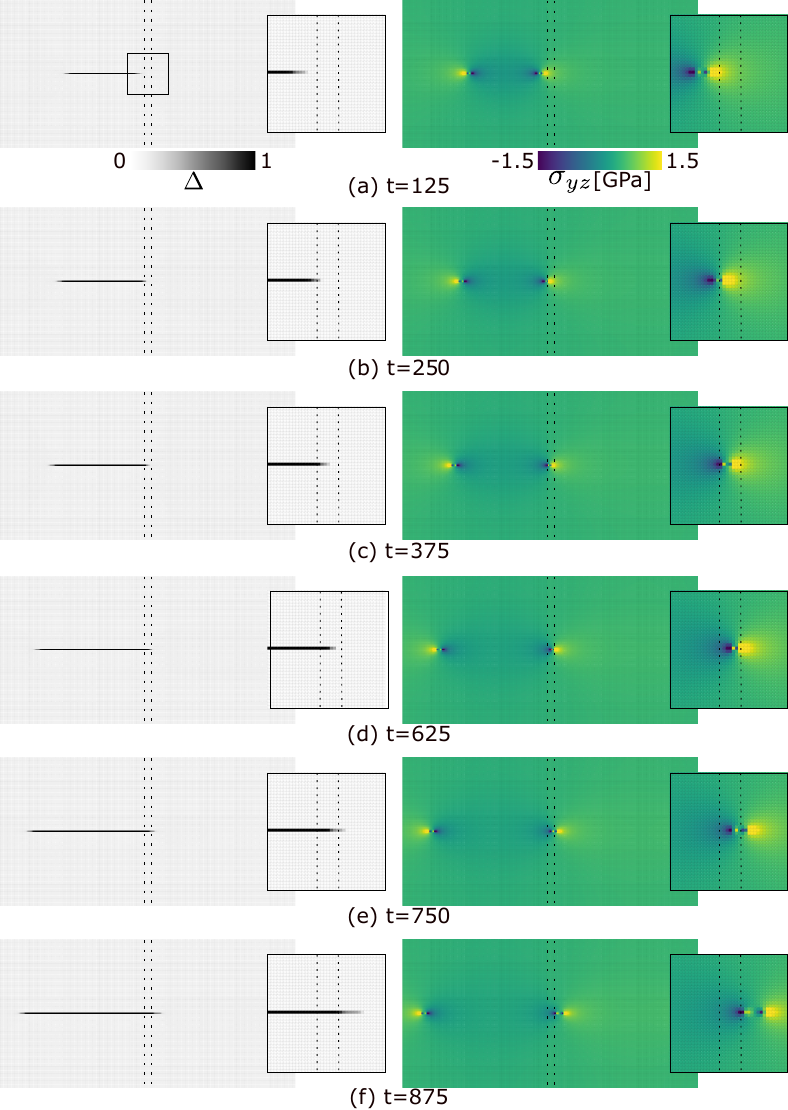}
   \caption{Disregistry $\Delta$ and stress component $\sigma_{yz}$ of a LAGB with GB energy $\gamma_{\mathsmaller{GB}}=1003$ mJ/m$^2$ under an applied stress of $\sigma_{yz}^{app} = 500$ MPa at t=125, 250, 375, 625, 750 and 875 (Case C1). The inclusion representing the GB is represented in the middle of the domain using dashed lines and the detailed view is shown at (a) t=125.}
   \label{fig:disevol-lagb-1003-500}
\end{figure}

\begin{figure}[ht!]
   \centering\leavevmode
   \includegraphics[width=135mm]{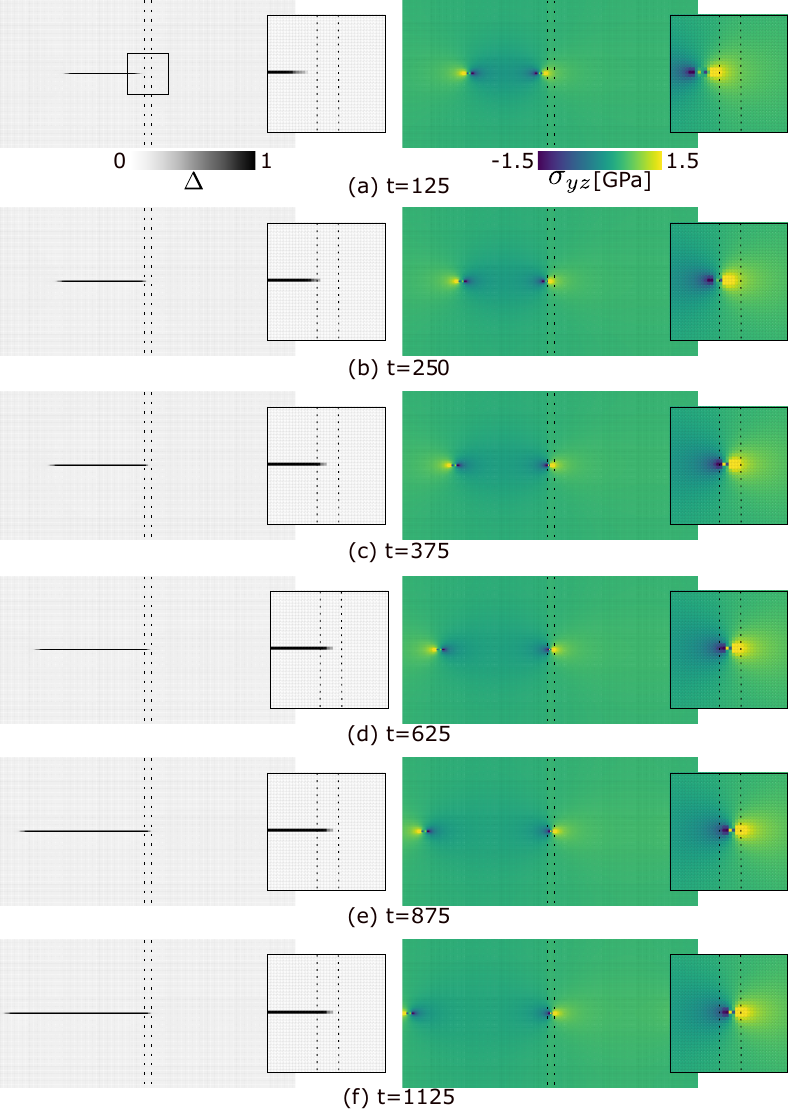}
   \caption{Disregistry $\Delta$ and stress component $\sigma_{yz}$ of a LAGB with GB energy $\gamma_{\mathsmaller{GB}}=1228$ mJ/m$^2$ under an applied stress of $\sigma_{yz}^{app} = 500$ MPa at t=125, 250, 375, 650, 875 and 1125 (Case C2). The inclusion representing the GB is represented in the middle of the domain using dashed lines and the detailed view is shown at (a) t=125.}
   \label{fig:disevol-lagb-1228-500}
\end{figure}

\begin{figure}[ht!]
   \centering\leavevmode
   \includegraphics[width=135mm]{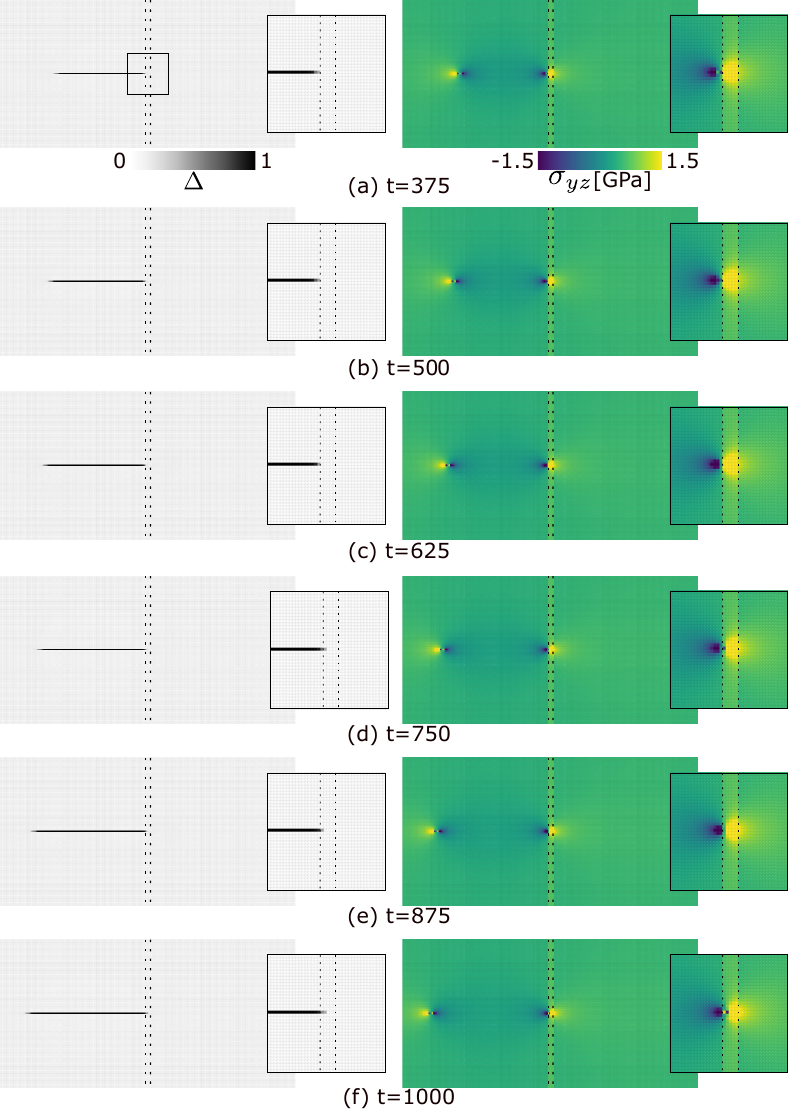}
   \caption{Disregistry $\Delta$ and stress component $\sigma_{yz}$ of a HAGB with GB energy $\gamma_{\mathsmaller{GB}}=847$ mJ/m$^2$ under an applied stress of $\sigma_{yz}^{app} = 500$ MPa at t=375, 500, 625, 750, 857 and 1000 (Case C3). The inclusion representing the GB is represented in the middle of the domain using dashed lines and the detailed view is shown at (a) t=375.}
   \label{fig:disevol-hagb-847-500}
\end{figure}

\begin{figure}[ht!]
   \centering\leavevmode
   \includegraphics[width=135mm]{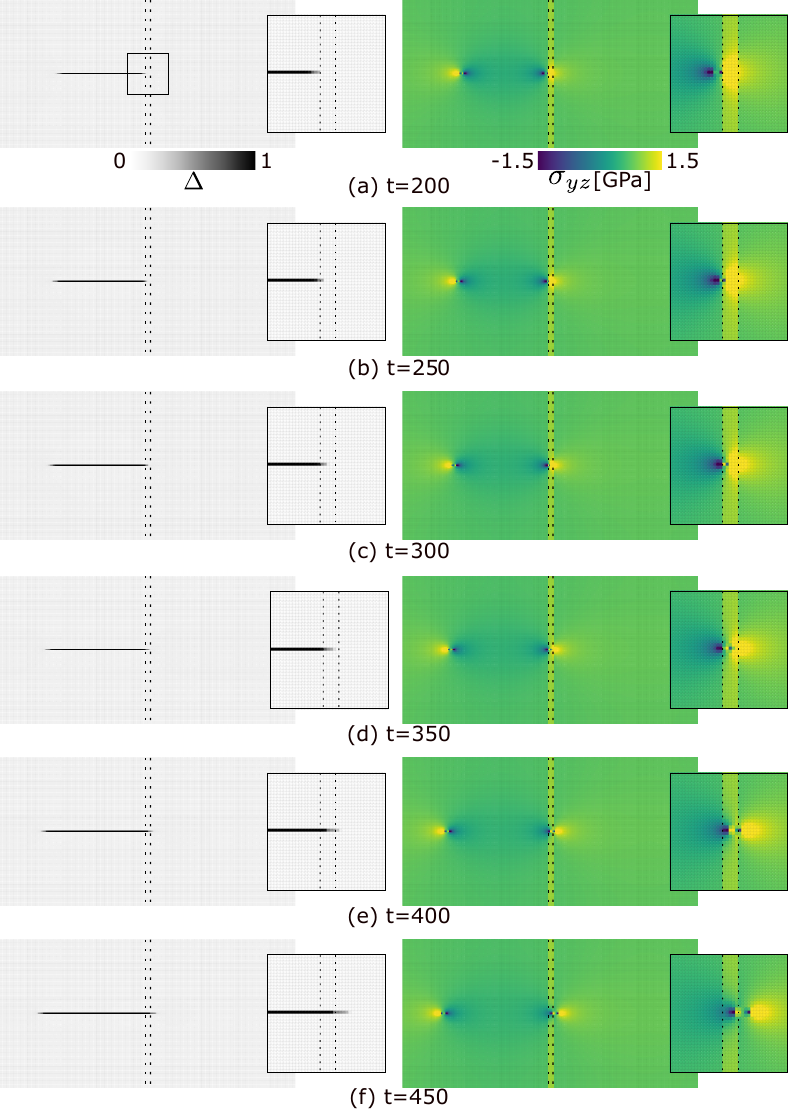}
   \caption{Disregistry $\Delta$ and stress component $\sigma_{yz}$ of a HAGB with GB energy $\gamma_{\mathsmaller{GB}}=847$ mJ/m$^2$ under an applied stress of $\sigma_{yz}^{app} = 750$ MPa at t=200, 250, 300, 350, 400 and 450 (Case C4). The inclusion representing the GB is represented in the middle of the domain using dashed lines and the detailed view is shown at (a) t=200.}
   \label{fig:disevol-hagb-847-750}
\end{figure}

Figures~\ref{fig:disevol-lagb-1003-500}, \ref{fig:disevol-lagb-1228-500}, \ref{fig:disevol-hagb-847-500} and \ref{fig:disevol-hagb-847-750} show the change of the disregistry and stress evolution at different time steps for cases C1, C2, C3 and C4. The disregistry is computed as \citep{shen2004incorporation}
\begin{equation}
    \Delta = \sum_{i=1}^{3} \zeta_i s_i \cdot s_{ini},
\end{equation}
where $s_{ini}$ is the Burgers vector direction of the initial perfect dislocation. The dislocation transmits in cases C1 and C4, and it is pinned in cases C2 and C3 as the dislocation stays blocked inside (Figure~\ref{fig:disevol-lagb-1228-500}) or next (Figure~\ref{fig:disevol-hagb-847-500}) to the GB region. Three main observations need to be highlighted, (i) transmission is easier for LAGB which is in accordance with MD simulations and experimental observations \citep{nieto2024assessment, adams2019atomistic}, (ii) the change in GB structure between C1 and C2 is reflected in the increase of GBE, which results in an increased resistance to the dislocation, causing it to be pinned, and (iii) the change in the interaction between cases C3 and C4 from being pinned to transmitted is a direct effect of the applied stress. Also, in C4 the dislocation evolves faster due to the higher applied stress of 750 MPa. 

\clearpage

\begin{figure}[ht!]
   \centering\leavevmode
   \includegraphics[width=90mm]{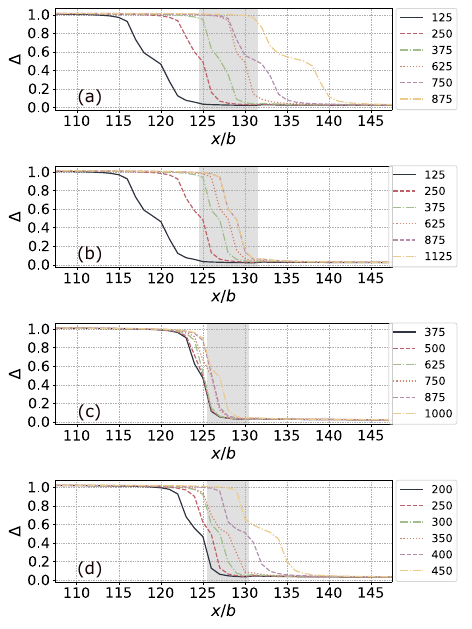}
   \caption{Disregistry evolution for a (a) LAGB with GB energy (GBE) $\gamma_{\mathsmaller{GB}}=1003$ mJ/m$^2$ under an applied stress of $\sigma_{yz}^{app} = 500$ MPa (Case C1), (b) LAGB with GBE $\gamma_{\mathsmaller{GB}}=1228$ mJ/m$^2$ under an applied stress of $\sigma_{yz}^{app} = 500$ MPa (Case C2), (c) HAGB with GBE $\gamma_{\mathsmaller{GB}}=847$ mJ/m$^2$ under an applied stress of $\sigma_{yz}^{app} = 500$ MPa (Case C3) and (d) HAGB with GBE $\gamma_{\mathsmaller{GB}}=847$ mJ/m$^2$ under an applied stress of $\sigma_{yz}^{app} = 750$ MPa (Case C4) at different time steps. The inclusion representing the GB is highlighted in the middle of the Figure in the gray shaded region.}
   \label{fig:deltaevol}
\end{figure}

\begin{figure}[ht!]
   \centering\leavevmode
   \includegraphics[width=160mm]{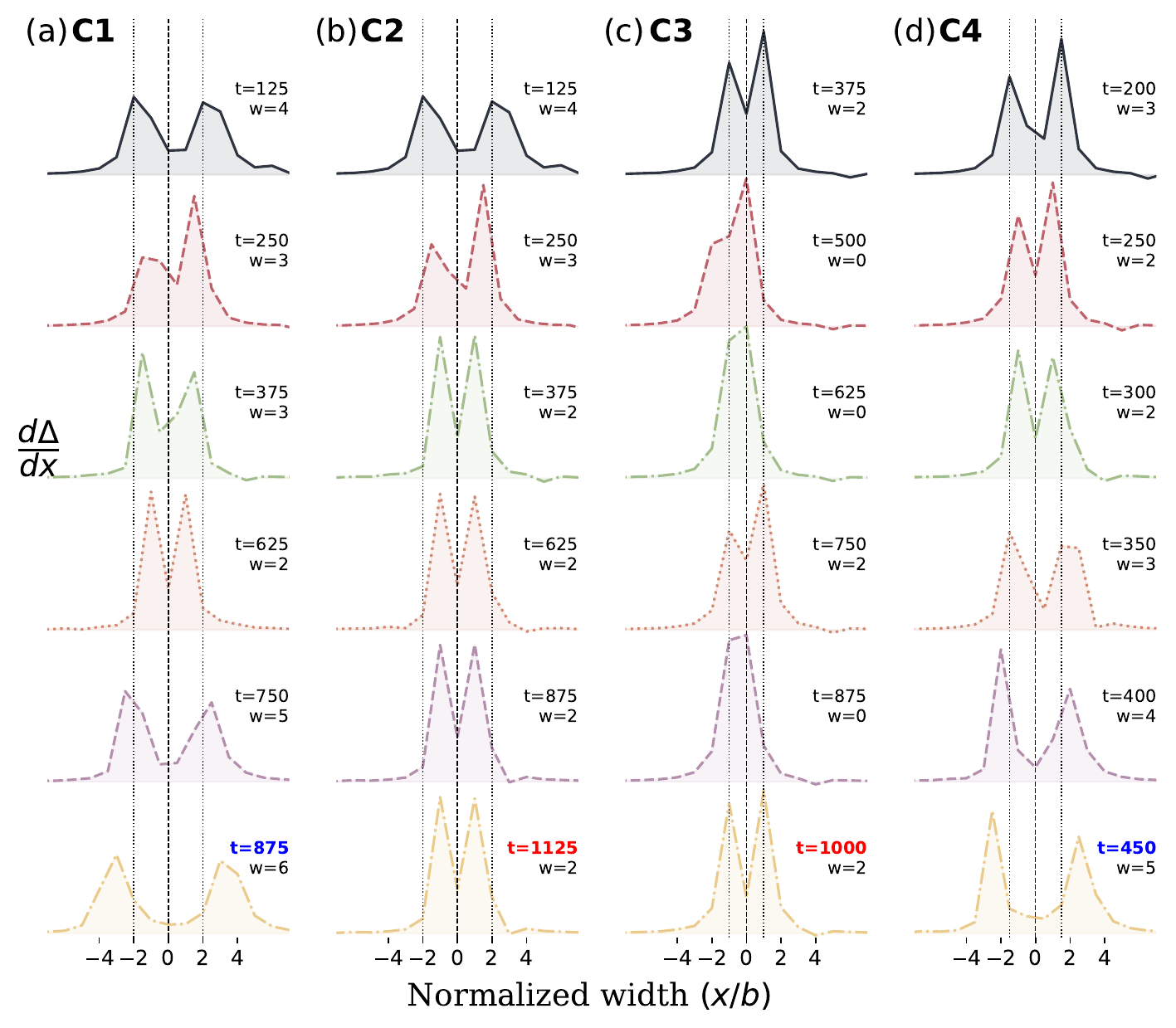}
   \caption{Disregistry derivative evolution for (a) case C1 LAGB with GB energy (GBE) $\gamma_{\mathsmaller{GB}}=1003$ mJ/m$^2$ under an applied stress of $\sigma_{yz}^{app} = 500$ MPa, (b) case C2 LAGB with GBE $\gamma_{\mathsmaller{GB}}=1228$ mJ/m$^2$ under an applied stress of $\sigma_{yz}^{app} = 500$ MPa, (c) case C3 HAGB with GBE $\gamma_{\mathsmaller{GB}}=847$ mJ/m$^2$ under an applied stress of $\sigma_{yz}^{app} = 500$ MPa and (d) case C4 HAGB with GBE $\gamma_{\mathsmaller{GB}}=847$ mJ/m$^2$ under an applied stress of $\sigma_{yz}^{app} = 750$ MPa at different time steps. The middle line represents the center of the two peaks and, the right and left lines represent the initial position of the peaks in the first time step shown in the top of the figure. The two times in blue indicate when the dislocation has transferred and the times in red indicate the times where the dislocation is pinned or absorbed.}
   \label{fig:ddeltaevol}
\end{figure}

The dislocation disregistry change during interaction with the GB is shown in Figure~\ref{fig:deltaevol}. Moreover, the change in the stacking fault width is clearly shown by plotting its derivative along the slip direction, see Figure~\ref{fig:ddeltaevol}. These detailed views of the disregistry are shown at the same time steps displayed in Figures~\ref{fig:disevol-lagb-1003-500}, \ref{fig:disevol-lagb-1228-500}, \ref{fig:disevol-hagb-847-500} and \ref{fig:disevol-hagb-847-750}. The initial value of the stacking fault width in the grains is 4 or 5 grid points as shown in Figure~\ref{fig:ddeltaevol} at t=125, corresponding to $\sim$1 nm, which is in agreement with previous studies \citep{jin2006interaction}. For the cases where the dislocation transmits across the GB, there is a difference in the stacking fault width during the interaction before and after leaving the GB region. In case C1, the stacking fault width decreases from 4 grid points to 3 grid points before the leading partial enters the GB region, as soon as both the trailing and leading partial enter the GB the stacking fault width decreases to 2 grid points and just before full transmission, the leading partial leaves the GB region generating an increase of the stacking fault width to 6 grid points see Figures~\ref{fig:disevol-lagb-1003-500}, \ref{fig:deltaevol}a and \ref{fig:ddeltaevol}. In case C4, the stacking fault width decreases to 2 grid points (Figure~\ref{fig:disevol-hagb-847-750}b), just before full transmission the stacking fault width is 5 grid points and comes back to its initial shape (Figure~\ref{fig:disevol-hagb-847-750}). Something similar happens in the cases that get pinned. In case C2, the dislocation gets pinned with the trailing and leading partials being blocked inside the GB region. Also, the stacking fault width changes slightly to 2 grid points as shown in Figures~\ref{fig:deltaevol}b and \ref{fig:ddeltaevol}. In case C3, as the dislocation approaches the GB and gets pinned the stacking fault width decreases to half of its original width from 4 to 2 grid points, see Figures~\ref{fig:deltaevol}c and \ref{fig:ddeltaevol}. As the perturbation to the lattice energy is higher in C3 than C2, the partial dislocations constricts to almost a perfect dislocation before the leading partial dislocation slightly enters the GB region. Due to this difference, the interaction in case C2 may be categorized as absorption instead of pinning. Note that absorption is a challenging interaction to model as the interaction can have different scenarios, e.g., dislocations can glide along the GB and re-emit elsewhere, or the dislocation can change the GB structure leading to nucleation of new dislocations at locations along the GB \citep{jin2006interaction, PhysRevMaterials.8.063604, liang2019slip, wang2015atomistic}. However, modeling absorption and subsequent dislocation glide along the GB plane is not available in the present PFDD model, but could be considered by adding existing glide and cross slip PFDD extensions \citep{fey2022phase, wise2025cross}. 

In summary, the variation in the stacking fault width is higher in C3 and C4 as shown in Figures~\ref{fig:deltaevol} and \ref{fig:ddeltaevol}. As $\Psi^{\mathsmaller{GB}}_{latt}$ is added to the total energy, it acts as an additional attractive force between the two partial dislocations which explains the change in the disregistry. Hence, the difference between the stacking fault width in C1, C2, C3 and C4 is proportional to $f_{\mathsmaller{GB}}$ despite of the higher applied stress in C4. In the Supplementary Material, $f_{\mathsmaller{GB}}$ is plotted against averaged strain gradient metrics showing good correlation with both, net strain gradient and gradient on uniaxial strain metrics computed using MS \citep{PhysRevMaterials.8.123605}. Similar changes in the stacking fault width have been reported in MD simulations by \cite{jin2006interaction} (cf. Figure 2), where the partial dislocation is constricted to an almost perfect dislocation at the GB before transmission or absorption of the partial dislocation, and by \cite{PhysRevMaterials.8.063604} (cf. Figure 2), where the partial dislocation stacking fault width reduces to almost half of its original width when the dislocation is pinned.

The remaining simulations are presented in the Supplementary Material, Supplementary Figures~\ref{fig:disevol-lagb-751-250}-\ref{fig:disevol-hagb-998-750} show the disregistry and stress evolution, and they are in accordance with the MD reactions as reported in Table~\ref{tab:GBfactor}. Another important observation is that the constriction of the stacking fault width is proportional to the applied stress. This can be seen in the HAGB cases with a GBE of 938 and 998 mJ/m$^2$. When the applied stress is 250 MPa, the partial dislocations get pinned at the GB, and the stacking fault width is about 3 grid points (See Supplementary Figures~\ref{fig:disevol-hagb-938-250} and \ref{fig:disevol-hagb-998-250}). When the applied stress is increased to 500 MPa, the partials constrict to almost a perfect dislocation (See Supplementary Figures~\ref{fig:disevol-hagb-938-500} and \ref{fig:disevol-hagb-998-500}). At 750 MPa, the LAGB with the highest value of GBE (1228 mJ/m$^2$) transmits as shown in Supplementary Figure~\ref{fig:disevol-lagb-1228-750} while the HAGB with highest GBE (998 mJ/m$^2$) is pinned as shown in Supplementary Figure~\ref{fig:disevol-hagb-998-750}.  Even if the GBE of the LAGB is higher, the perturbation to the lattice energy is higher for the HAGB resulting from a combination of the fitting constant, GBE, and the elastic constants. This is clearly evident from the values of $f_{GB}$ in Table~\ref{tab:GBfactor}. This higher perturbation to the lattice energy results in a higher barrier to dislocation transmission in the GB region and consequently higher stress for transmission. Note that the elastic constants of the HAGB and the stress at the GB region is higher than the LAGB, this may affect the reaction. A similar observation was also reported in a recent MD study by \cite{wu2026transmission}.



\clearpage

\subsection{Considering multiple GBs}

\begin{figure}[ht]
   \centering\leavevmode
   \includegraphics[width=135mm]{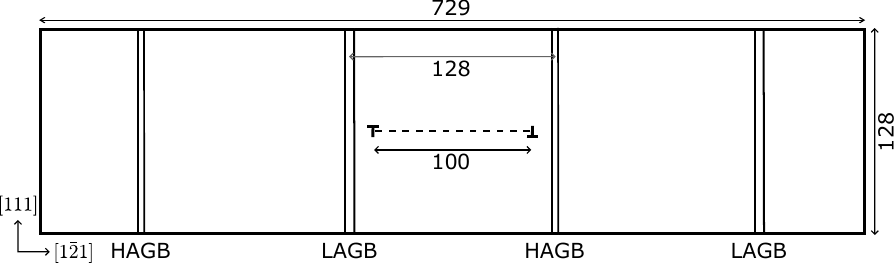}
   \caption{Large scale simulation setup showing the position of the dislocation dipole (in number of grid points) and inclusions representing the GBs in a polycrystal (4 crystals). Distance between the center of two neighboring GBs is 128 grid points. The domain size is 729x2x128. The dislocation dipole is centered in the vertical direction, z ($[111]$), the GB is centered in the horizontal direction, x ($[1\bar{2}1]$), and the width of the LAGB and HAGB is 7 and 5 grid points, respectively.}
   \label{fig:HetDislocGBs}
\end{figure}

One of the main advantages of mesoscopic simulations is length and time scales can be larger than those considered in MD simulations. In this section, an example of a larger scale simulation is shown where multiple GBs are placed along the domain. The domain size is 729x2x128 grid points with a grid size corresponding to a Burgers vector and periodic boundary conditions in the three directions. Two LAGBs and two HAGBs are placed along the $[1\bar{2}1]$ direction, with properties corresponding to the minimum energy state in Table~\ref{tab:GBprops}. A screw dislocation dipole is placed in the center of the domain, with a distance between the dislocations of 100 grid points and slip direction $[10\bar{1}]$, see Figure~\ref{fig:HetDislocGBs}. The domain is subjected to a shear stress $\sigma_{yz}^{app}=424.48$ MPa $(0.0105*\mu_{grain})$.

\begin{figure}[ht!]
   \centering\leavevmode
   \includegraphics[width=110mm]{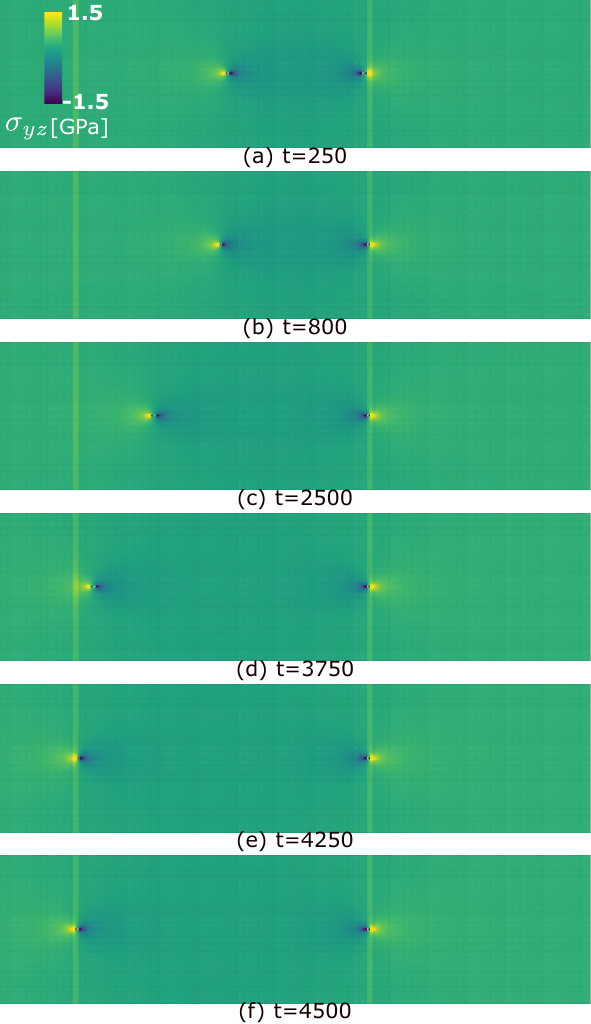}
   \caption{Screw dislocation evolution and interaction with multiple GBs under an applied stress of $\sigma_{yz}^{app} = 424.48$ MPa. Stress component $\sigma_{yz}$ at (a) t=250 (b) 800, (c) 2500, (d) 3750, (e) 4250 and (d) 4500.}
   \label{fig:disevol-multiple-gbs}
\end{figure}

Figure~\ref{fig:disevol-multiple-gbs} shows the stress field as the dislocations evolve. Each dislocation dissociates into two partials and starts to evolve towards a GB. Due to the different GB properties, the partials interact differently with the GBs, the positive partial dislocations (right) get pinned at the HAGB and the negative partial dislocations (left) get transmitted to the adjacent grain at the early stage of the simulation (See Figure~\ref{fig:disevol-multiple-gbs}a, b and c). The negative partial dislocations continue to evolve to the left (Figure~\ref{fig:disevol-multiple-gbs}d) until it interacts with the next HAGB, constricts and get pinned (Figure~\ref{fig:disevol-multiple-gbs}e and f). This simulation shows the versatility of the new coupled PFDD approach; it allows the model to consider the interaction of dislocations with multiple GBs and can be used to study interaction of more complex GB networks as those present in polycrystals. 

\clearpage

\section{Conclusion}
\label{sec:conclusion}

A new PFDD formulation that couples phase field and micromechanical solvers is presented. This new formulation allows for the study of heterogeneous materials with different stiffness tensors by solving the Lippmann-Schwinger equation. The total strain obtained from the Lippmann-Schwinger equation is one of the terms in the elastic energy density used within the time-dependent Ginzburg-Landau kinetic equation. 

In this work, GB-dislocation interactions are studied by representing GBs as inclusions with a specific width obtained from MS simulations \citep{PhysRevMaterials.8.123605}. Two additional parameters are added to describe GBs, (i) an elastic energy which is considered by computing a different GB elastic stiffness tensor and (ii) a lattice energy that represents the effect of the disordered atomic structure within the GB region. Two different groups of GBs are studied including their minimum energy and metastable states: LAGB with GB plane (1,1,6) and misorientation angle of 27$^\circ$ and HAGB with GB plane (5,5,2) and misorientation angle of 178$^\circ$. The GB-dislocations interactions predicted using PFDD show excellent agreement with the MD results. However, these results depend on the calibration parameter $C_{\mathsmaller{GB}}$, which varies as a function of the misorientation. We presented an analysis on the stacking fault width, the initial width and reduction of the width during the interaction which shows good agreement with MD simulations \citep{jin2006interaction, PhysRevMaterials.8.063604}. Finally, by adding an additional lattice energy for the GB region proportional to $f_{\mathsmaller{GB}}$, which correlated with strain gradient metrics derived from atomistic simulations, we can account for the dependency of GB-dislocation interaction on the GB energy or GB structure. 

In contrast to the sharp interface method \citep{zeng2016phase, ma2022dislocation}, this new methodology opens the door to model more general GB-dislocation interactions. Future work will be focused on the improvement of the GB description and the inclusion of other interactions such as GB absorption and glide of the dislocation along the GB. Also, other GB-dislocation interactions such as reflection should be considered. Finally, after the model is calibrated using MD data, it is possible to study GB-dislocation interactions in bigger systems such as polycrystals or potentially GBs with varying structure along their length.

\section*{Supplementary material}

The supplementary material presents the basic scheme algorithm, details of the coupled formulation, the remaining PFDD simulations and plots of strain functional descriptors as a function of $f_{GB}$

\section*{Acknowledgments}

AH, BM, and NM acknowledge support from the Physics and Engineering Models (PEM) Materials project within the Advanced Simulation and Computing (ASC) Program at LANL (ASC-PEM-Materials Project). AM and NM also acknowledges funding from the LDRD-DR “Investigating How Material’s Interfaces and Dislocations Affect Strength” (iMIDAS) under Grant No. 20210036DR (Abigail Hunter, PI and Saryu J. Fensin, Co-PI) and the LDRD-PRD project 20220814PRD4: ``Grain Boundary Characterization from Diffractograms by Physics-Informed Machine Learning'' for the initial MD calculations and methodology. BM also acknowledges Ian Chesser and Ian Wise for their constructive comments and discussions during the development of the work. LA-UR-25-26159 version 2.

\section*{Data availability}

The data cannot be made publicly available upon publication because they are not available in a format that is sufficiently accessible or reusable by other researchers. The data that support the findings of this study are available upon reasonable request from the authors.

\section*{Conflicts of interest}

The authors declare no conflicts of interest.


\bibliographystyle{apalike}

\bibliography{main}

\newpage

\setcounter{page}{1}

\setcounter{affn}{0}
\setcounter{section}{0}
\resetTitleCounters
\graphicspath{{figures/}}

\makeatletter
\let\@title\@empty
\makeatother

\title{Supplementary material}

\newcommand{\beginsupplement}{
    \setcounter{table}{0}
    \renewcommand{\thetable}{\arabic{table}}
    \renewcommand{\tablename}{Supplementary Table}
    \setcounter{figure}{0}
    \renewcommand{\thefigure}{\arabic{figure}}
    \renewcommand{\figurename}{Supplementary Figure}
    \setcounter{equation}{0}
    \renewcommand{\theequation}{S.\arabic{equation}}
}

\makeatletter
\def\ps@pprintTitle{%
  \let\@oddhead\@empty
  \let\@evenhead\@empty
  \def\@oddfoot{\reset@font\hfil\thepage\hfil}
  \let\@evenfoot\@oddfoot
}
\makeatother

\beginsupplement

\maketitle

\section{Micromechanics solver: Basic scheme}

\begin{algorithm}[H]
    \caption{Basic scheme - Nonlinear \citep{moulinec1994fast}}
    \begin{algorithmic}[1]
        \REQUIRE $C^m, \Delta C(x), tol, E, \varepsilon^p$ 
        \RETURN $\varepsilon_{ij}$
        \IF{t=0}
            \STATE $\varepsilon_{mn}^{i=0} = E_{mn}$,
        \ELSE
            \STATE $\varepsilon_{mn}^{i=0} = \varepsilon_{mn}^{t-\Delta t}$, 
        \ENDIF
        \WHILE{$e > tol$}
            \STATE $ \sigma_{ij}^i(x) = ( C_{ijmn}^0 + \Delta C_{ijmn}(x) ) ( E_{mn} + \varepsilon_{mn}^i(x)  - \varepsilon_{mn}^p(x) )  $
            \STATE $ \tau_{ij}^i(x) = \sigma_{ij}^i(x) - C_{ijmn}^0 \varepsilon_{mn}^i(x)   $
            \STATE $\hat{\tau}_{ij}^i(\xi) = F| \tau_{ij}(x) |$
            \STATE $ \hat{\varepsilon}_{kl}^{i+1}(\xi) = - \Gamma_{klij} \hat{\tau}_{ij}^i(\xi) $
            \STATE $\varepsilon_{kl}^{i+1}(x) = F^{-1}| \hat{\varepsilon}_{ij}^{i+1}(\xi) |  $ 
            \STATE $e = \dfrac{\|\varepsilon^t - \varepsilon^{t-\Delta t}\|}{\|\varepsilon^{t-\Delta t}\|}$
        \ENDWHILE
    \end{algorithmic}
    \label{alg:basic}
\end{algorithm}
The fourth-order Green operator is given by 
\begin{equation}
    \Gamma_{khij} = \dfrac{1}{4} ( \xi_j \xi_k ( C^0_{nhim} \xi_m \xi_n )^{-1} + \xi_j \xi_h ( C^0_{knim} \xi_m \xi_n )^{-1} + \xi_i \xi_k ( C^0_{nhmj} \xi_m \xi_n )^{-1}  + \xi_k \xi_j ( C^0_{mhin} \xi_m \xi_n )^{-1}) \, ,
\end{equation}
where $\xi$ is the wave number vector. In the particular case of isotropic elasticity it may redefined as
\begin{equation}
    \Gamma_{khij} = \dfrac{1}{4 \mu^0 |\xi|^2 }( \delta_{ki} \xi_h \xi_j + \delta_{hi} \xi_k \xi_j + \delta_{kj} \xi_h \xi_i + \delta_{hj} \xi_k \xi_i ) - \dfrac{\lambda^0 + \mu^0}{\mu^0 (\lambda^0 + 2\mu^0) } \dfrac{\xi_i \xi_j \xi_k \xi_h}{|\xi|^4}.
\end{equation}

\section{Details of the coupled formulation}
\label{suppSec:details}
This sections aims to show some of the steps that were not shown in Section~\ref{sec:methods} to get the derivative of the elastic energy density with respect to the order parameter and the macroscopic strain in Equation~\ref{eqn:homstraincoupled}.

The elastic energy is computed in the real space using Equations~\ref{eqn:straindecomposition}, \ref{eqn:LS} and \ref{eqn:elastreal}. The derivative of the elastic energy density with respect to the order parameter is computed as 
\begin{equation}
    \dfrac{\partial \psi_{elas}}{\partial \zeta^{\alpha}} = C_{rskl}(x) \dfrac{\partial \varepsilon_{rs}^{el} (x)}{\partial \zeta^{\alpha}} \varepsilon_{kl}^{el} (x) \, ,
    \label{eqn:elasticEnergyDensDerivative}
\end{equation}
with
\begin{equation}
    \dfrac{\partial \varepsilon_{rs}^{el} (x)}{\partial \zeta^{\alpha}} = \dfrac{\partial}{\partial \zeta^{\alpha}} \left( E_{rs} - \Gamma^0_{rspq} \star \tau_{pq} - \varepsilon^p_{rs} \right) \, .
    \label{eqn:elasticStrainDerivative}
\end{equation}
Assuming that $\dfrac{\partial E_{rs}}{\partial \zeta^{\alpha}} = 0$ and $\dfrac{\partial \varepsilon_{rs}}{\partial \zeta^{\alpha}} = 0$, Equation~\ref{eqn:elasticStrainDerivative} is deifned as
\begin{equation}
\begin{split}
    \dfrac{\partial \varepsilon_{rs}^{el} (x)}{\partial \zeta^{\alpha}} = & - \Gamma^0_{rspq} \star \dfrac{\partial \tau_{pq} }{\partial \zeta^{\alpha}} - \dfrac{\partial \varepsilon^p_{rs}}{\partial \zeta^{\alpha}} \\
    = & - \Gamma^0_{rspq} \star \dfrac{\partial}{\partial \zeta^{\alpha}} \left( \Delta  C_{pqmn}(x)  ( \varepsilon_{mn} + E_{mn} ) - C_{pqmn}(x) \varepsilon^p_{mn} \right) \\
    & - \dfrac{\partial}{\partial \zeta^{\alpha}} \left( \sum^{N}_{\alpha=1} \dfrac{b^{\alpha}}{d^{\alpha}} [s^{\alpha} \otimes n^{\alpha} ]_{rs} \zeta^{\alpha} (x) \right) \\ 
    = & \Gamma^0_{rspq} \star \dfrac{\partial}{\partial \zeta^{\alpha}} \left( C_{pqmn}(x) \sum^{N}_{\alpha=1} \dfrac{b^{\alpha}}{d^{\alpha}} [s^{\alpha} \otimes n^{\alpha} ]_{mn} \zeta^{\alpha} (x) \right) - \dfrac{b^{\alpha}}{d^{\alpha}} [s^{\alpha} \otimes n^{\alpha} ]_{rs} \\
    = & \Gamma^0_{rspq} \star \left( C_{pqmn}(x) \dfrac{b^{\alpha}}{d^{\alpha}} [s^{\alpha} \otimes n^{\alpha} ]_{mn} \right) - \dfrac{b^{\alpha}}{d^{\alpha}} [s^{\alpha} \otimes n^{\alpha} ]_{rs} \, .
\end{split}
\end{equation}
Thus, Equation~\ref{eqn:elasticEnergyDensDerivative} is redefined as
\begin{equation}
    \dfrac{\partial \psi_{elas}}{\partial \zeta^{\alpha}} = C_{rskl}(x) \left( \Gamma^0_{rspq} \star \left( C_{pqmn}(x) \dfrac{b^{\alpha}}{d^{\alpha}} [s^{\alpha} \otimes n^{\alpha} ]_{mn} \right) - \dfrac{b^{\alpha}}{d^{\alpha}} [s^{\alpha} \otimes n^{\alpha} ]_{rs} \right) \left( E_{kl} + \varepsilon_{kl} - \varepsilon^p_{kl} \right) \, ,
    \label{eqn:elasticEnergyDensDerivative2}
\end{equation}
where the second term is solved in the Fourier space
\begin{equation} 
    \dfrac{\partial \psi_{elas}}{\partial \zeta^{\alpha}} = C_{rskl}(x) \left( \mathcal{F}^{-1} \left[ \hat{\Gamma}^0_{rspq} \hat{C}_{pqmn}(\xi) \dfrac{b^{\alpha}}{d^{\alpha}} [s^{\alpha} \otimes n^{\alpha} ]_{mn} \right] - \dfrac{b^{\alpha}}{d^{\alpha}} [s^{\alpha} \otimes n^{\alpha} ]_{rs} \right)  \left( E_{kl} + \varepsilon_{kl} - \varepsilon^p_{kl} \right) \, .
\end{equation}
In a compact form, it can be expressed as
\begin{equation}
    \dfrac{\partial \psi_{elas}}{\partial \zeta^{\alpha}} = [FF^{\alpha}_{kl}(x)] \left( E_{kl} +  \varepsilon_{kl} - \varepsilon^p_{kl} \right) \, ,
\end{equation}
with the interaction matrix defined as
\begin{equation}
    [FF^{\alpha}_{kl}(x)] = C_{rskl}(x) \left( \mathcal{F}^{-1} \left[  \hat{\Gamma}^0_{rspq} \hat{C}_{pqmn}(\xi) \dfrac{b^{\alpha}}{d^{\alpha}} [s^{\alpha} \otimes n^{\alpha} ]_{mn} \right] - \dfrac{b^{\alpha}}{d^{\alpha}} [s^{\alpha} \otimes n^{\alpha} ]_{rs} \right) \, .
\end{equation} 
Finally, Equation~\ref{eqn:homstraincoupled} is obtained through the derivation of the strain energy with respect to the average strain. The strain energy is defined as:
\begin{equation}
\begin{split}
    \Psi_{str} = & \dfrac{1}{2} \int_v C_{ijkl} (x) \varepsilon_{ij}^e \varepsilon_{kl}^e dV - \sigma_{ij}^{app} E_{ij} V \\
    = & \dfrac{1}{2} \int_v C_{ijkl} (x) (\varepsilon_{ij} + E_{ij}) (\varepsilon_{kl} + E_{kl}) dV - \int_v C_{ijkl} (x) (\varepsilon_{ij} + E_{ij}) \varepsilon_{kl}^p dV \\
    & + \dfrac{1}{2} \int_v C_{ijkl} (x) \varepsilon_{ij}^p \varepsilon_{kl}^p dV - \sigma_{ij}^{app} E_{ij} V,
\end{split}
\end{equation}
making the derivative of the strain energy with respect to the average strain equal to zero and using the major symmetry of the elastic modulus:
\begin{equation}
    \dfrac{\partial \Psi_{str}}{\partial E_{ij}} = \int_v C_{ijkl} (x) (\varepsilon_{kl} + E_{kl}) dV - \int_v C_{ijkl} (x) \varepsilon_{kl}^p - V \sigma_{ij}^{app} = 0 \, .
\end{equation}
Rearranging the terms and dividing by the volume gives
\begin{equation}
    E_{kl} \dfrac{1}{V} \int_v C_{ijkl} (x) dV = - \dfrac{1}{V} \int_v \Delta C_{ijkl} (x) \varepsilon_{kl} dV + \dfrac{1}{V} \int_v C_{ijkl} (x) \varepsilon_{kl}^p dV + \sigma_{ij}^{app} \, .
\end{equation}
Note that by definition the term $ C_{ijkl}^0 \int_v \varepsilon_{kl} dV = 0$. Equaton~\ref{eqn:homstraincoupled} is obtained using the definitions of averaged quantities presented by the end of Section~\ref{sec:methods} ($ \big< S_{ijkl} \big> = \big< C_{ijkl} \big>^{-1} $, $ \big< C_{ijkl} \big> =  \frac{1}{V} \int_v C_{ijkl} (x) dV $, $\big< \sigma_{ij}^{p}  \big> = \frac{1}{V} \int_v C_{ijkl} (x) \varepsilon_{kl}^p dV$, and $ \big< \sigma_{ij} \big> =  \frac{1}{V} \int_v \Delta C_{ijkl} (x) \varepsilon_{kl} dV $).



\clearpage

\section{Phase field dislocation dynamics simulations}

In the main text only three cases were presented, the remaining simulations are shown in Supplementary Figures~\ref{fig:disevol-lagb-751-250}-\ref{fig:disevol-hagb-998-750}. 

\begin{figure}[ht!]
   \centering\leavevmode
   \includegraphics[width=135mm]{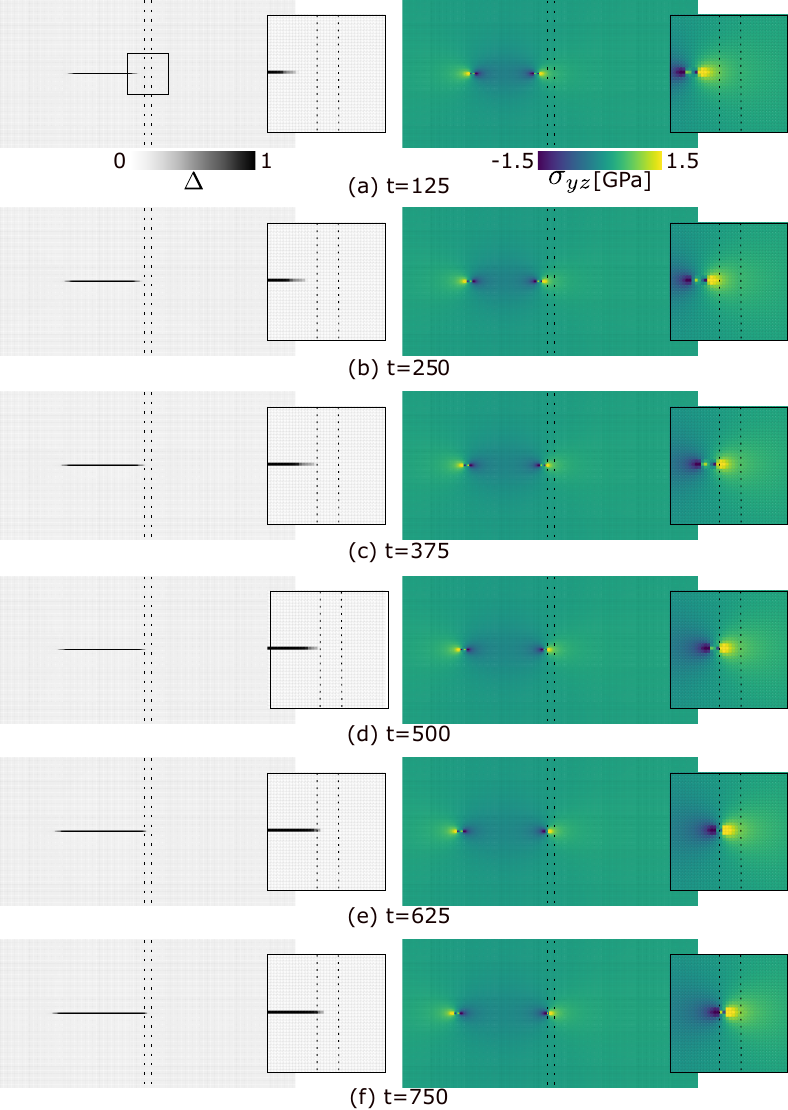}
   \caption{Disregistry $\Delta$ and stress component $\sigma_{yz}$ of a LAGB with GB energy $\gamma_{\mathsmaller{GB}}=751$ mJ/m$^2$ under an applied stress of $\sigma_{yz}^{app} = 750$ MPa at t=125, 250, 375, 500, 625 and 750. The inclusion representing the GB is represented in the middle of the domain using dashed lines and the detailed view is shown at (a) t=125.}
   \label{fig:disevol-lagb-751-250}
\end{figure}

\begin{figure}[ht!]
   \centering\leavevmode
   \includegraphics[width=135mm]{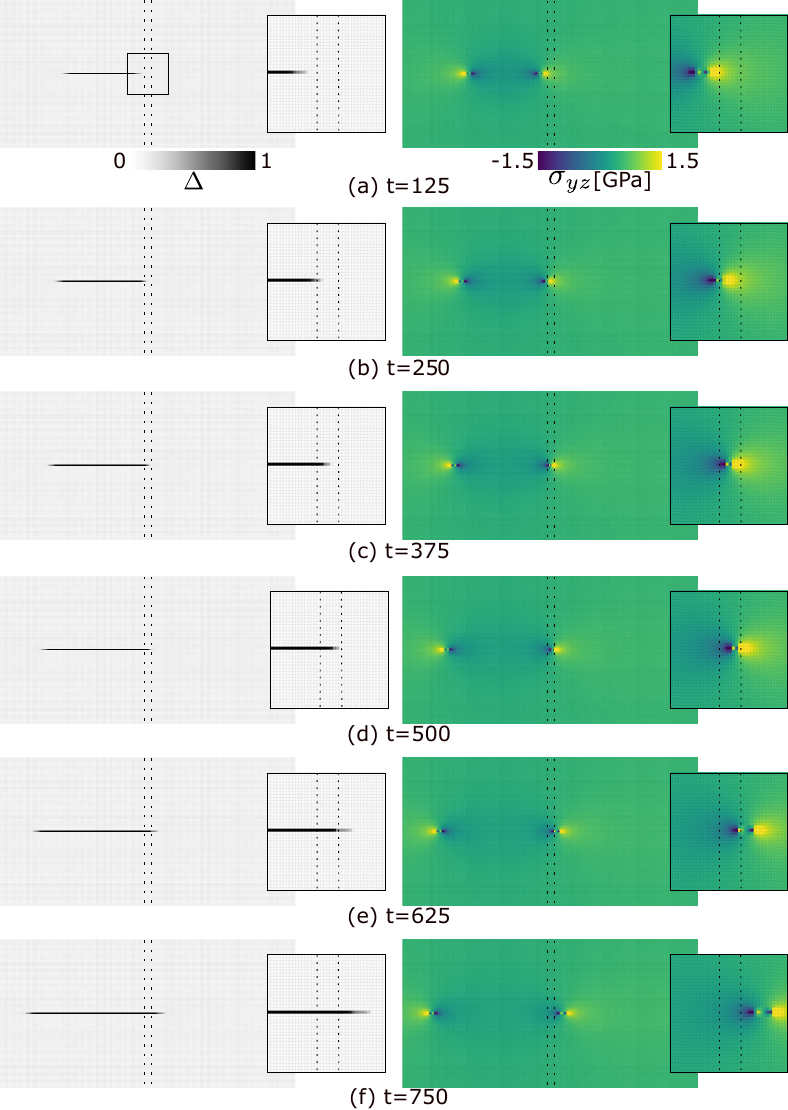}
   \caption{Disregistry $\Delta$ and stress component $\sigma_{yz}$ of a LAGB with GB energy $\gamma_{\mathsmaller{GB}}=751$ mJ/m$^2$ under an applied stress of $\sigma_{yz}^{app} = 750$ MPa at t=125, 250, 375, 500, 625 and 750. The inclusion representing the GB is represented in the middle of the domain using dashed lines and the detailed view is shown at (a) t=125.}
   \label{fig:disevol-lagb-751-500} 
\end{figure}

\begin{figure}[ht!]
   \centering\leavevmode
   \includegraphics[width=135mm]{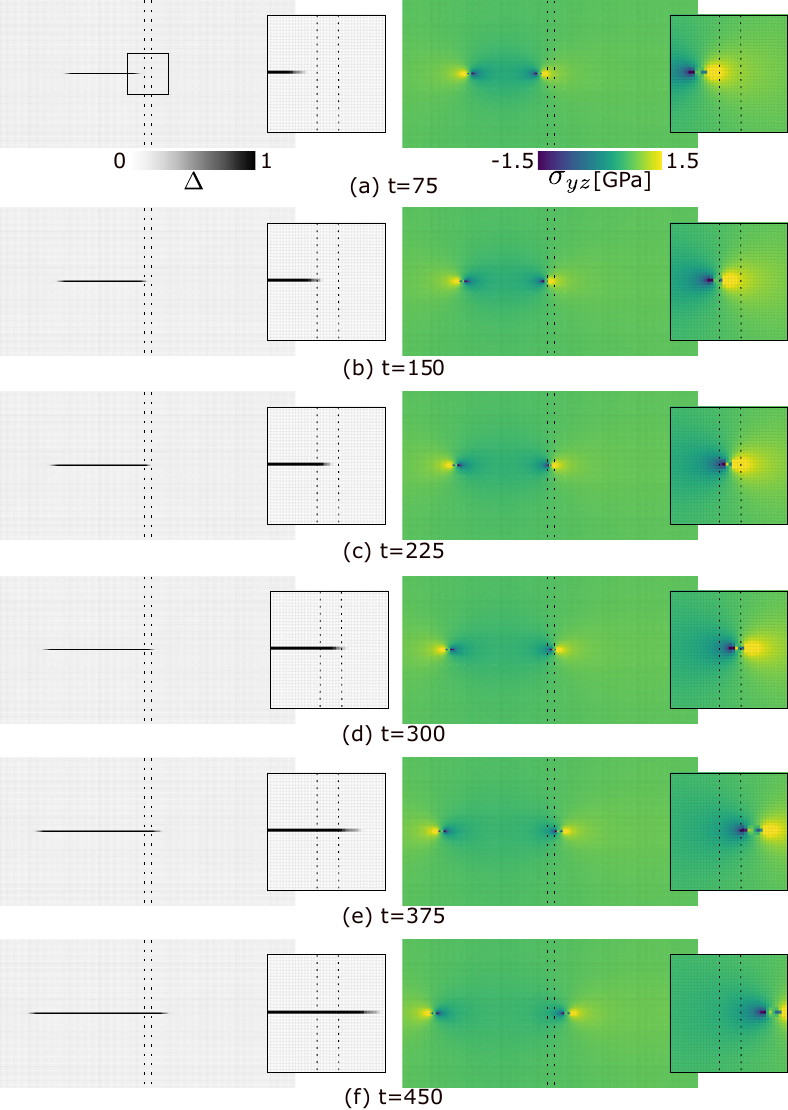}
   \caption{Disregistry $\Delta$ and stress component $\sigma_{yz}$ of a LAGB with GB energy $\gamma_{\mathsmaller{GB}}=751$ mJ/m$^2$ under an applied stress of $\sigma_{yz}^{app} = 750$ MPa at t=75, 150, 225, 300, 375 and 450. The inclusion representing the GB is represented in the middle of the domain using dashed lines and the detailed view is shown at (a) t=75.}
   \label{fig:disevol-lagb-751-750}
\end{figure}

\begin{figure}[ht!]
   \centering\leavevmode
   \includegraphics[width=135mm]{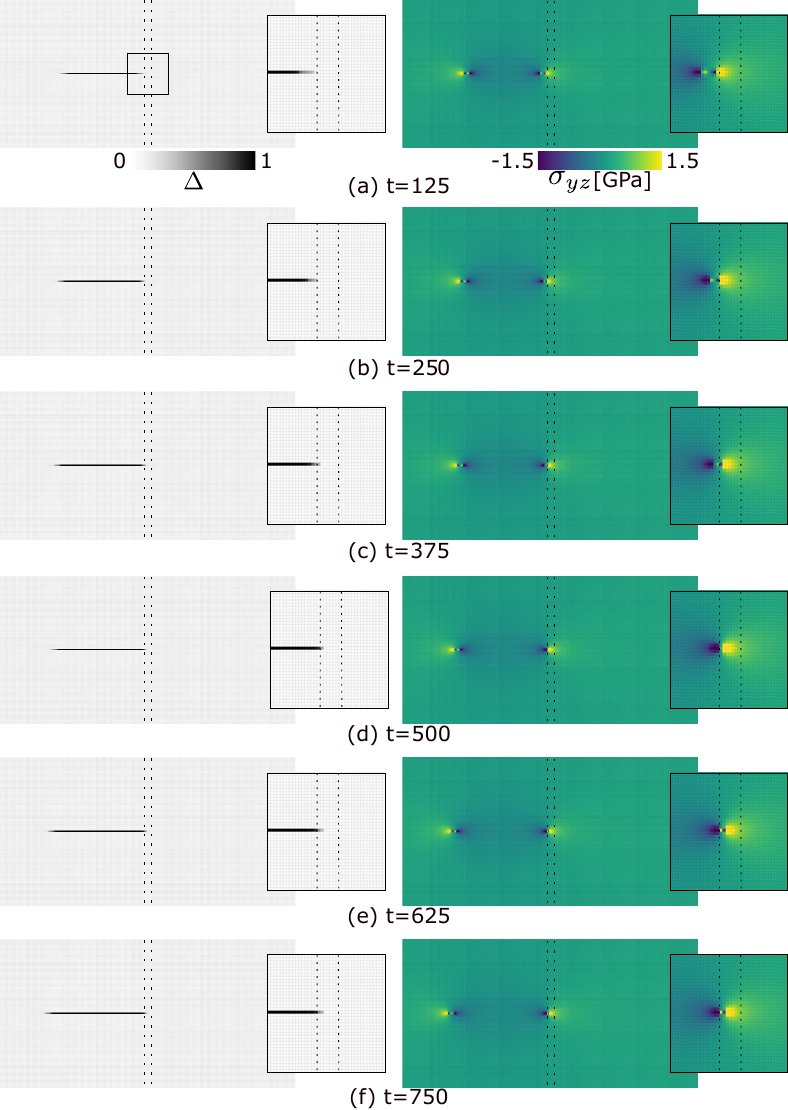}
   \caption{Disregistry $\Delta$ and stress component $\sigma_{yz}$ of a LAGB with GB energy $\gamma_{\mathsmaller{GB}}=1003$ mJ/m$^2$ under an applied stress of $\sigma_{yz}^{app} = 250$ MPa at t=125, 250, 375, 500, 625 and 750. The inclusion representing the GB is represented in the middle of the domain using dashed lines and the detailed view is shown at (a) t=125.}
   \label{fig:disevol-lagb-1003-250}
\end{figure}

\begin{figure}[ht!]
   \centering\leavevmode
   \includegraphics[width=135mm]{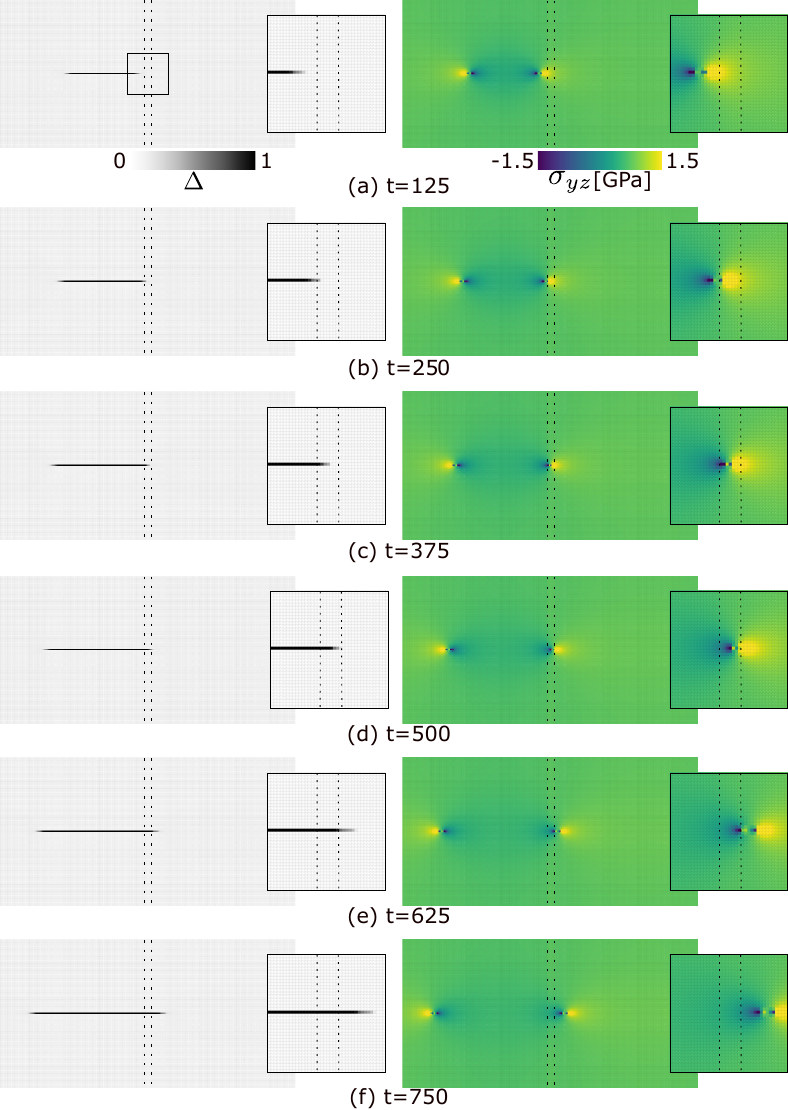}
   \caption{Disregistry $\Delta$ and stress component $\sigma_{yz}$ of a LAGB with GB energy $\gamma_{\mathsmaller{GB}}=1003$ mJ/m$^2$ under an applied stress of $\sigma_{yz}^{app} = 750$ MPa at t=125, 250, 375, 500, 625 and 750. The inclusion representing the GB is represented in the middle of the domain using dashed lines and the detailed view is shown at (a) t=125.}
   \label{fig:disevol-lagb-1003-750}
\end{figure}

\begin{figure}[ht!]
   \centering\leavevmode
   \includegraphics[width=135mm]{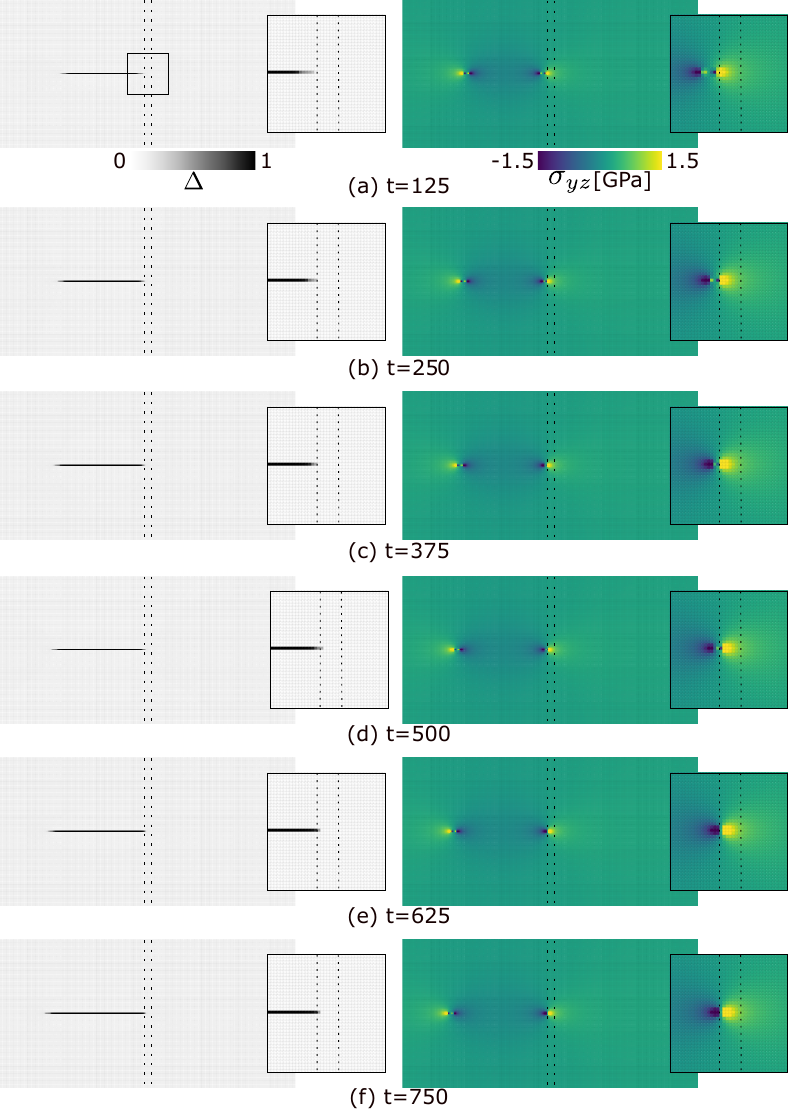}
   \caption{Disregistry $\Delta$ and stress component $\sigma_{yz}$ of a LAGB with GB energy $\gamma_{\mathsmaller{GB}}=1228$ mJ/m$^2$ under an applied stress of $\sigma_{yz}^{app} = 250$ MPa at t=125, 250, 375, 500, 625 and 750. The inclusion representing the GB is represented in the middle of the domain using dashed lines and the detailed view is shown at (a) t=125.}
   \label{fig:disevol-lagb-1228-250}
\end{figure}

\begin{figure}[ht!]
   \centering\leavevmode
   \includegraphics[width=135mm]{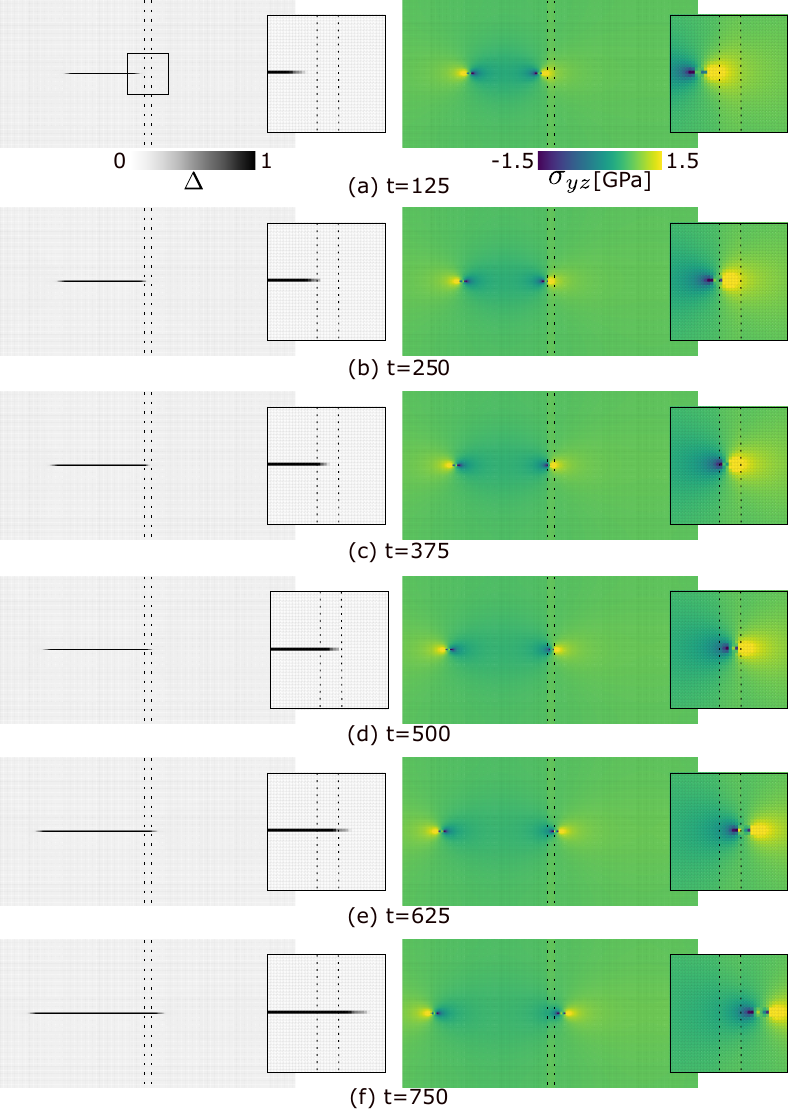}
   \caption{Disregistry $\Delta$ and stress component $\sigma_{yz}$ of a LAGB with GB energy $\gamma_{\mathsmaller{GB}}=1228$ mJ/m$^2$ under an applied stress of $\sigma_{yz}^{app} = 750$ MPa at t=125, 250, 375, 500, 625 and 750. The inclusion representing the GB is represented in the middle of the domain using dashed lines and the detailed view is shown at (a) t=125.}
   \label{fig:disevol-lagb-1228-750}
\end{figure}

\begin{figure}[ht!]
   \centering\leavevmode
   \includegraphics[width=135mm]{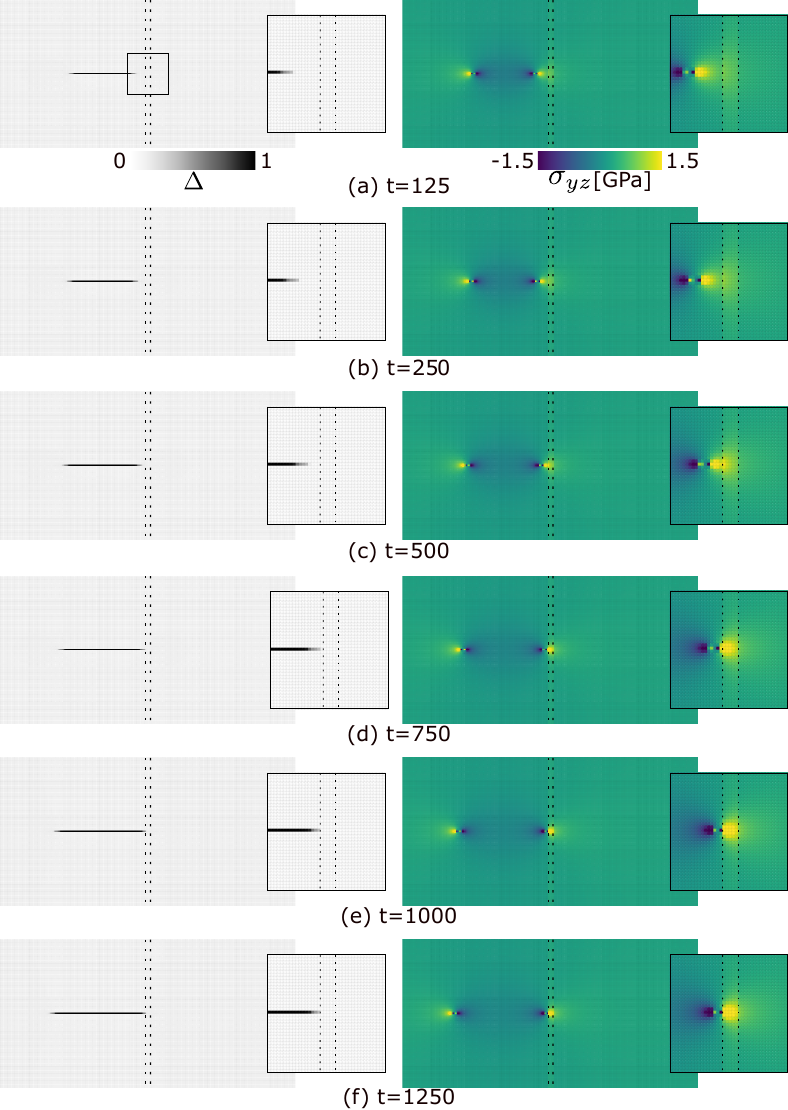}
   \caption{Disregistry $\Delta$ and stress component $\sigma_{yz}$ of a HAGB with GB energy $\gamma_{\mathsmaller{GB}}=847$ mJ/m$^2$ under an applied stress of $\sigma_{yz}^{app} = 250$ MPa at t=125, 250, 500, 750, 1000 and 1250. The inclusion representing the GB is represented in the middle of the domain using dashed lines and the detailed view is shown at (a) t=125.}
   \label{fig:disevol-hagb-847-250}
\end{figure}

\begin{figure}[ht!]
   \centering\leavevmode
   \includegraphics[width=135mm]{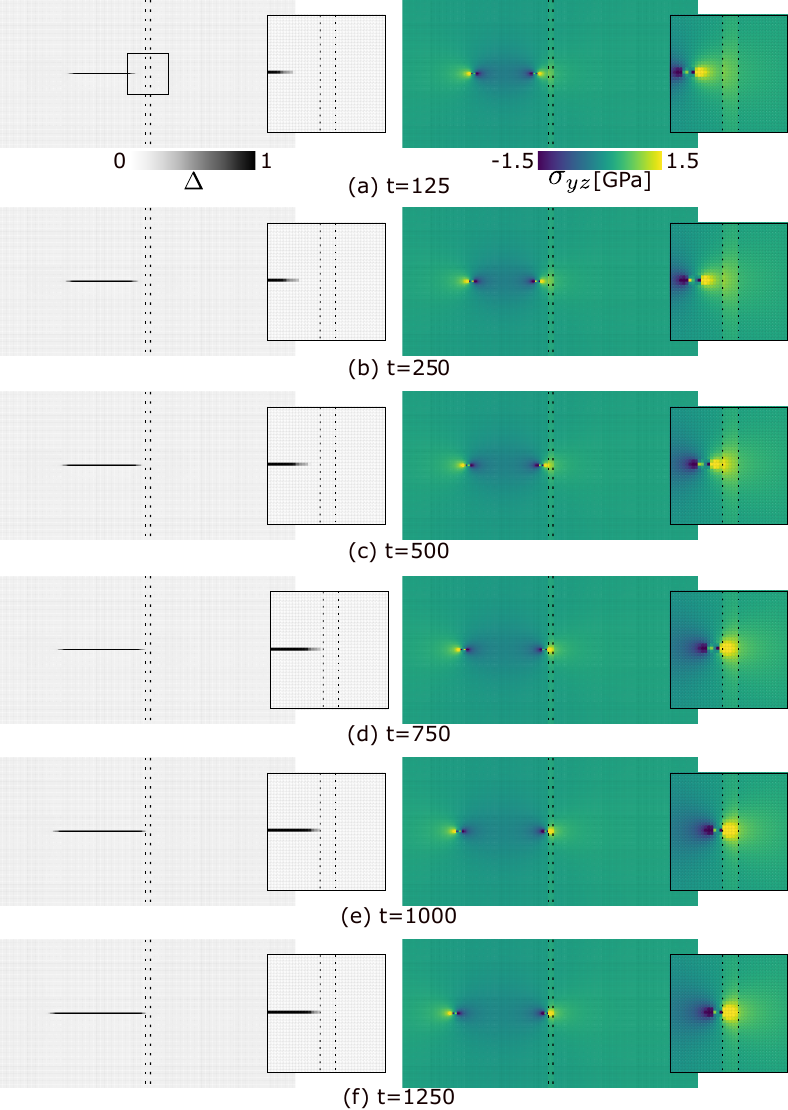}
   \caption{Disregistry $\Delta$ and stress component $\sigma_{yz}$ of a HAGB with GB energy $\gamma_{\mathsmaller{GB}}=938$ mJ/m$^2$ under an applied stress of $\sigma_{yz}^{app} = 250$ MPa at t=125, 250, 500, 750, 1000 and 1250. The inclusion representing the GB is represented in the middle of the domain using dashed lines and the detailed view is shown at (a) t=125.}
   \label{fig:disevol-hagb-938-250}
\end{figure}

\begin{figure}[ht!]
   \centering\leavevmode
   \includegraphics[width=135mm]{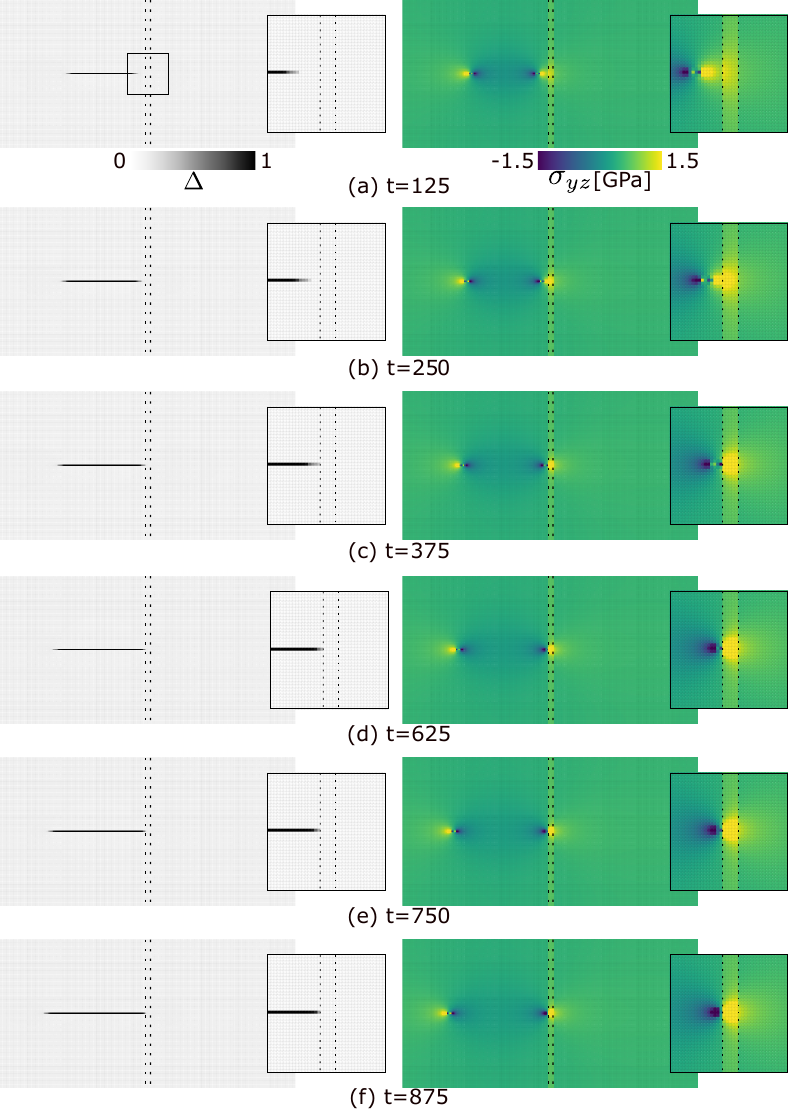}
   \caption{Disregistry $\Delta$ and stress component $\sigma_{yz}$ of a HAGB with GB energy $\gamma_{\mathsmaller{GB}}=938$ mJ/m$^2$ under an applied stress of $\sigma_{yz}^{app} = 500$ MPa at t=125, 250, 375, 625, 750 and 875. The inclusion representing the GB is represented in the middle of the domain using dashed lines and the detailed view is shown at (a) t=125.}
   \label{fig:disevol-hagb-938-500}
\end{figure}

\begin{figure}[ht!]
   \centering\leavevmode
   \includegraphics[width=135mm]{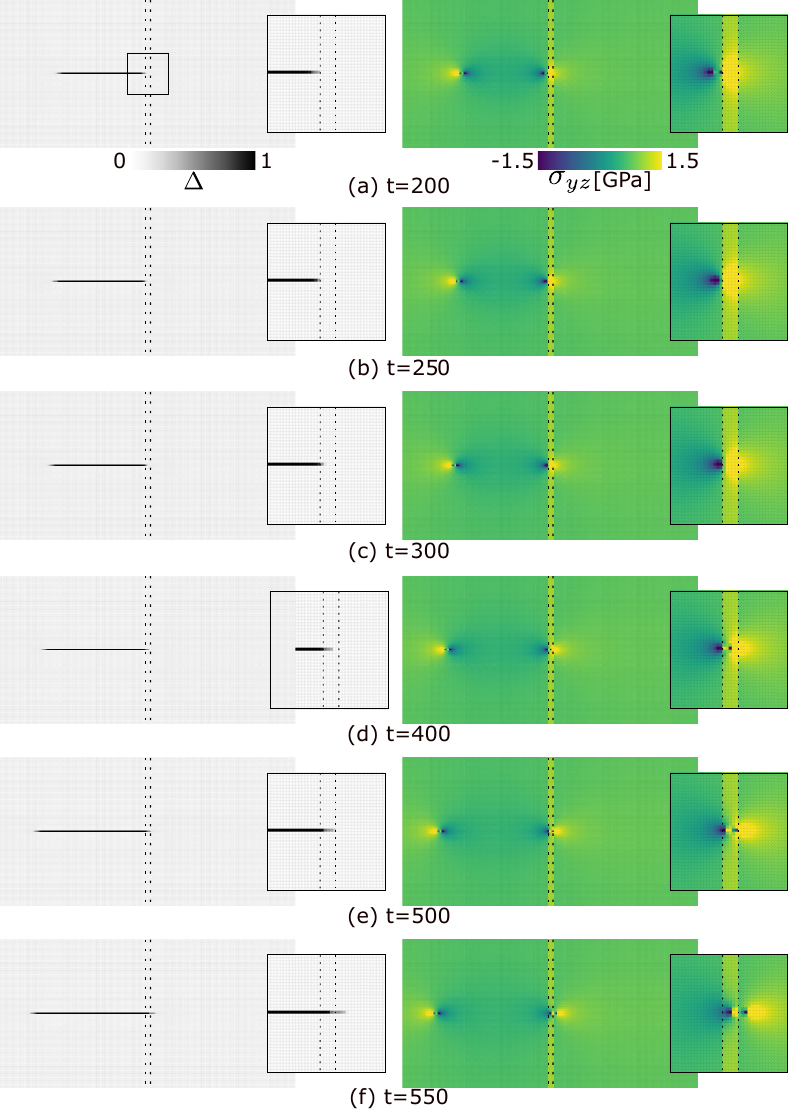}
   \caption{Disregistry $\Delta$ and stress component $\sigma_{yz}$ of a HAGB with GB energy $\gamma_{\mathsmaller{GB}}=938$ mJ/m$^2$ under an applied stress of $\sigma_{yz}^{app} = 750$ MPa at t=200, 250, 300, 400, 500 and 550. The inclusion representing the GB is represented in the middle of the domain using dashed lines and the detailed view is shown at (a) t=200.}
   \label{fig:disevol-hagb-938-750}
\end{figure}

\begin{figure}[ht!]
   \centering\leavevmode
   \includegraphics[width=135mm]{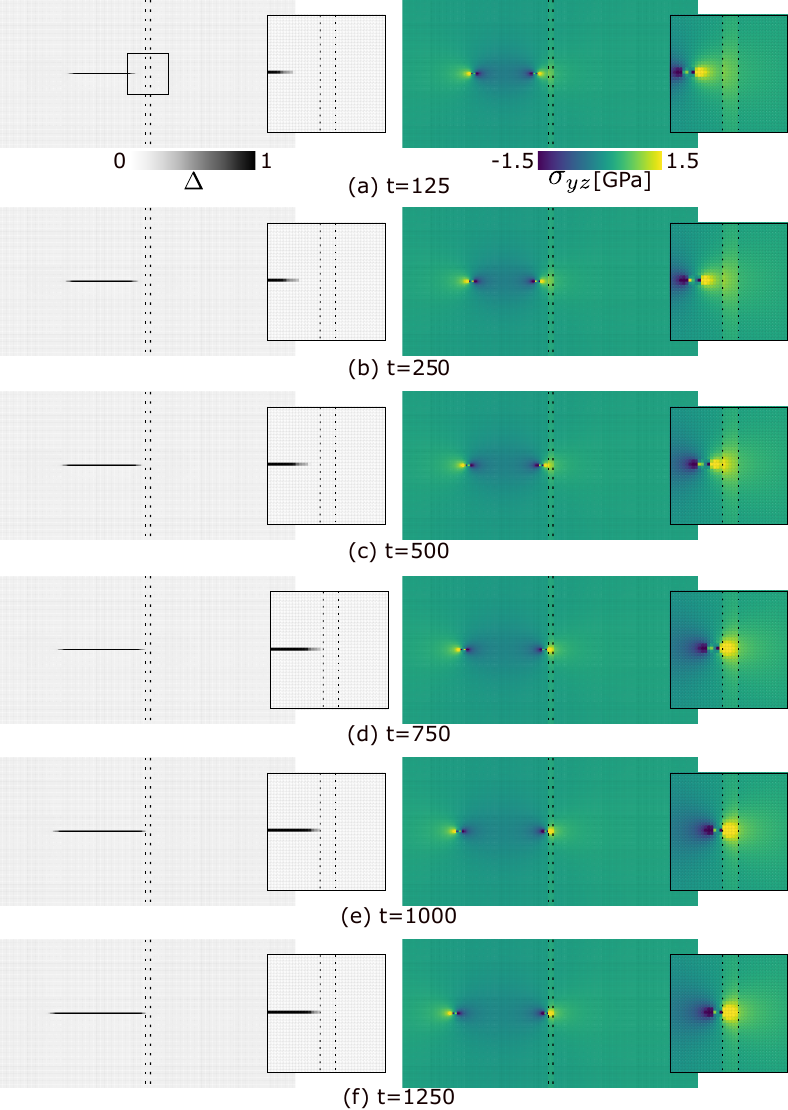}
   \caption{Disregistry $\Delta$ and stress component $\sigma_{yz}$ of a HAGB with GB energy $\gamma_{\mathsmaller{GB}}=998$ mJ/m$^2$ under an applied stress of $\sigma_{yz}^{app} = 250$ MPa at t=125, 250, 500, 750, 1000 and 1250. The inclusion representing the GB is represented in the middle of the domain using dashed lines and the detailed view is shown at (a) t=125.}
   \label{fig:disevol-hagb-998-250}
\end{figure}

\begin{figure}[ht!]
   \centering\leavevmode
   \includegraphics[width=135mm]{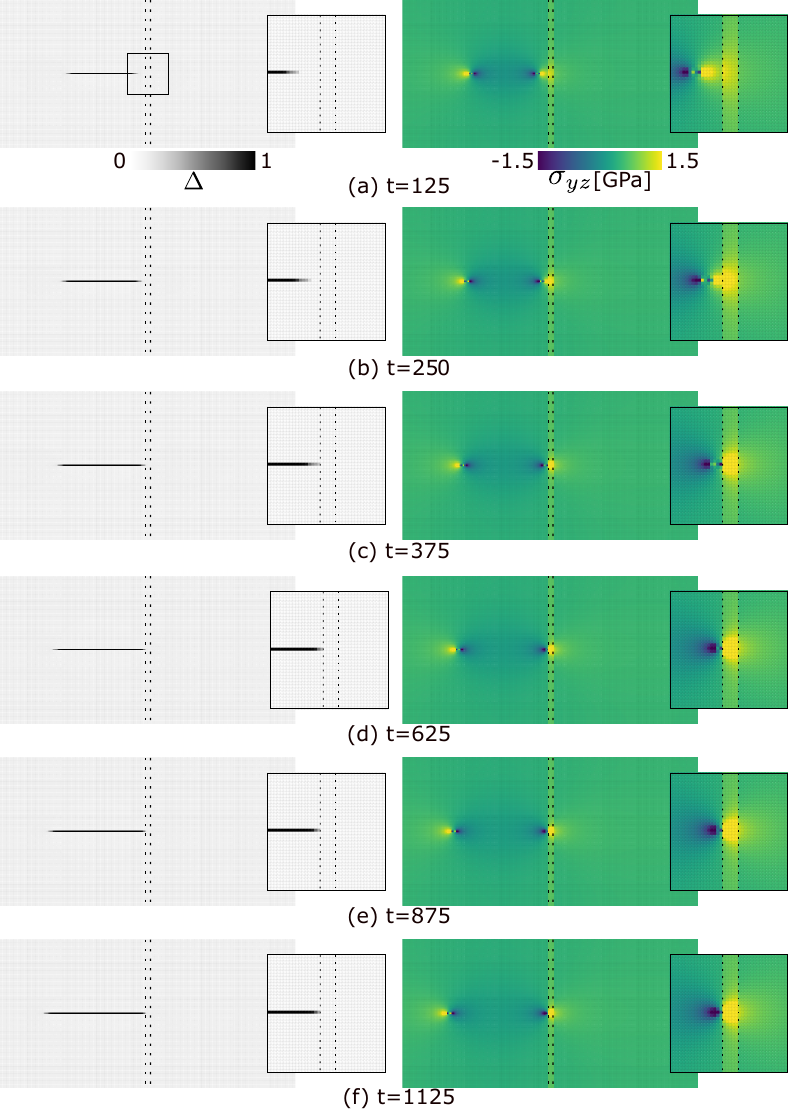}
   \caption{Disregistry $\Delta$ and stress component $\sigma_{yz}$ of a HAGB with GB energy $\gamma_{\mathsmaller{GB}}=998$ mJ/m$^2$ under an applied stress of $\sigma_{yz}^{app} = 500$ MPa at t=125, 250, 375, 625, 875, and 1125. The inclusion representing the GB is represented in the middle of the domain using dashed lines and the detailed view is shown at (a) t=125.}
   \label{fig:disevol-hagb-998-500}
\end{figure}

\begin{figure}[ht!]
   \centering\leavevmode
   \includegraphics[width=135mm]{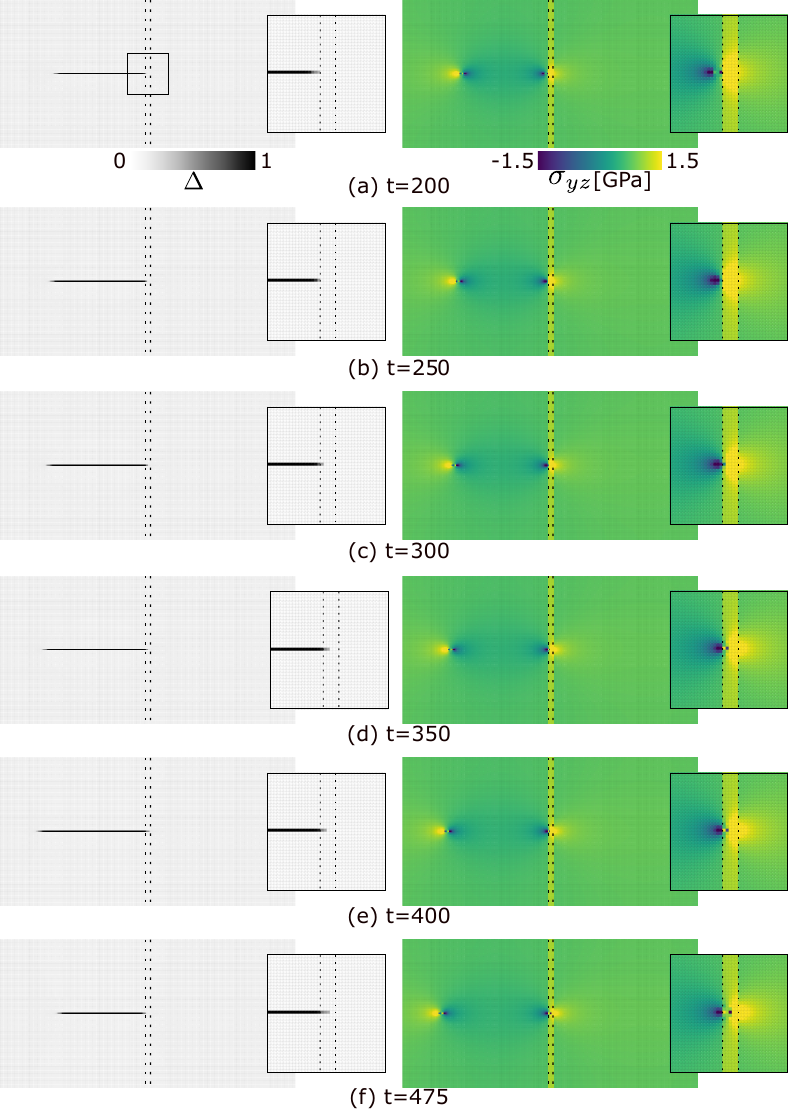}
   \caption{Disregistry $\Delta$ and stress component $\sigma_{yz}$ of a HAGB with GB energy $\gamma_{\mathsmaller{GB}}=998$ mJ/m$^2$ under an applied stress of $\sigma_{yz}^{app} = 750$ MPa at t=200, 250, 300, 350, 400 and 475. The inclusion representing the GB is represented in the middle of the domain using dashed lines and the detailed view is shown at (a) t=200.}
   \label{fig:disevol-hagb-998-750}
\end{figure}

\clearpage

\begin{figure}[ht!]
   \centering\leavevmode
   \includegraphics[width=135mm]{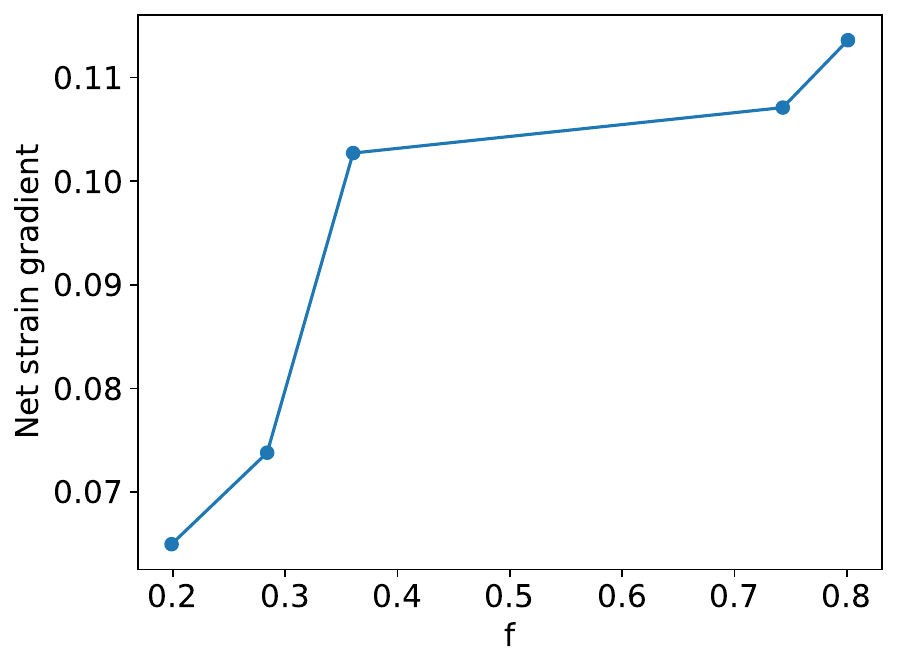}
   \caption{Strain functional descriptor for the net strain gradient as a function of the $f_{GB}$ values of the GBs considered in the study.}
   \label{fig:P3I0-smallGB}
\end{figure}

\begin{figure}[ht!]
   \centering\leavevmode
   \includegraphics[width=135mm]{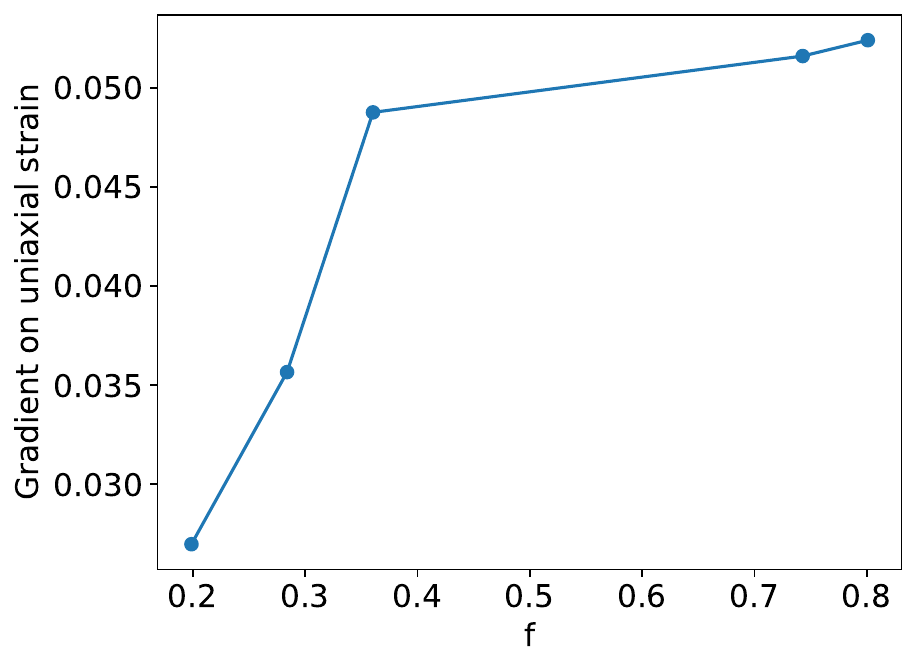}
   \caption{Strain functional descriptor for the gradient of uniaxial strain as a function of the $f_{GB}$ values of the GBs considered in the study.}
   \label{fig:P3I4-smallGB}
\end{figure}

Averaged strain-gradient metrics calculated for classification in \citep{PhysRevMaterials.8.123605} are plotted as a function of $f_{\mathsmaller{GB}}$ in Figures \ref{fig:P3I0-smallGB} and \ref{fig:P3I4-smallGB}. These were computed using the small atomistic simulation cells used for GB width and elastic stiffness calculations. The average values were computed over a region around the GB corresponding to the GB width. These show that the  $f_{\mathsmaller{GB}}$ are correlated with strain-gradient metrics. These GBs had the same macroscopic degrees of freedom and similar energy values to the GBs used in dislocation-GB interaction study \citep{dang2025dislocation}. 

\begin{figure}[ht!]
   \centering\leavevmode
   \includegraphics[width=135mm]{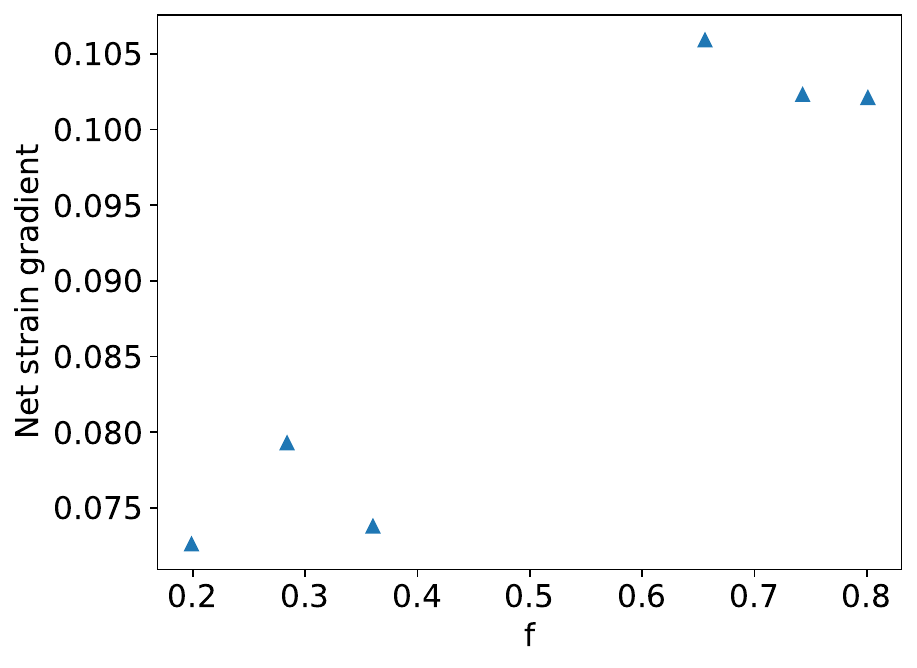}
   \caption{Strain functional descriptor for the net strain gradient as a function of the $f_{GB}$ values of the GBs considered in the study, computed using the large MD simulation cells.}
   \label{fig:P3I0-largeGB}
\end{figure}

\begin{figure}[ht!]
   \centering\leavevmode
   \includegraphics[width=135mm]{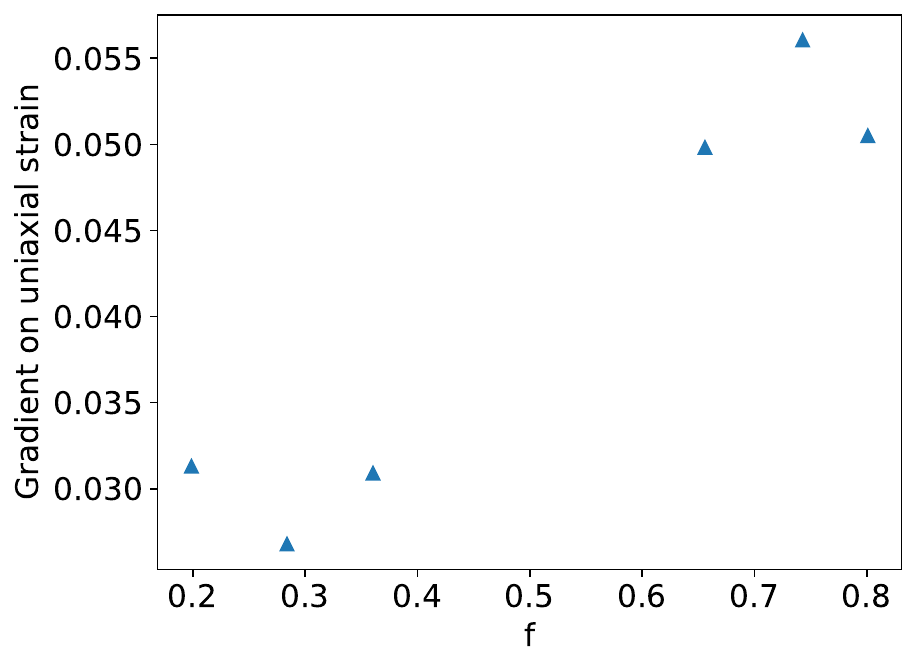}
   \caption{Strain functional descriptor for the gradient of uniaxial strain as a function of the $f_{GB}$ values of the GBs considered in the study, computed using the large MD simulation cells.}
   \label{fig:P3I4-largeGB}
\end{figure}

The data shows similar trends when computed from large simulation cells as shown in Figures \ref{fig:P3I0-largeGB} and \ref{fig:P3I4-largeGB}. In this case the averages were computed using all atoms in a region around the dislocation interaction site, defined by the GB width (normal to the GB plane) and 60 \AA (normal to the tilt axis). As the larger atomistic simulation cells are less constrained, the GB undergoes relaxation resulting in differences in energy values compared to the small cells. 

\clearpage

\end{document}